\documentclass[twocolumn,showpacs,preprintnumbers,amsmath,amssymb,prc,nofootinbib,final]{revtex4-1}
\usepackage{graphicx}% Include figure files
\usepackage{dcolumn}% Align table columns on decimal point
\usepackage{bm}% bold math
\usepackage{multirow}
\usepackage{color}
\usepackage{hyperref}

\usepackage{subfigure}
\usepackage{dcolumn}
\usepackage{natbib}
\usepackage[subfigure,caption2]{ccaption}
\usepackage{color}
\usepackage[normalem]{ulem}
\usepackage{soul,xcolor}

\include{comdefs}

\begin{document}

\title{Derivation of an optical potential for statically deformed rare-earth nuclei from a global spherical potential}

\author{G. P. A. Nobre}
\email[Corresponding author: ]{gnobre@bnl.gov}
\affiliation{National Nuclear Data Center, Brookhaven National Laboratory, Upton, NY 11973-5000, USA}
\author{A. Palumbo}
\affiliation{National Nuclear Data Center, Brookhaven National Laboratory, Upton, NY 11973-5000, USA}

\author{F. S. Dietrich}
\affiliation{P.O. Box 30423, Walnut Creek, CA, 94598, USA}
\author{M. Herman}
\affiliation{National Nuclear Data Center, Brookhaven National Laboratory, Upton, NY 11973-5000, USA}
\author{D. Brown}
\affiliation{National Nuclear Data Center, Brookhaven National Laboratory, Upton, NY 11973-5000, USA}
\author{S. Hoblit}
\affiliation{National Nuclear Data Center, Brookhaven National Laboratory, Upton, NY 11973-5000, USA}

\begin{abstract}
The coupled-channel theory is a natural way of treating nonelastic channels,
in particular those arising from collective excitations characterized by nuclear deformations.
A proper treatment of such excitations is often essential to the accurate description of experimental
nuclear-reaction data and to the prediction of a wide variety of scattering observables. Stimulated by
recent work substantiating the near validity of the adiabatic approximation in coupled-channel
calculations for scattering on statically deformed nuclei, we explore the possibility of
generalizing a global spherical optical model potential (OMP) to make it usable in coupled-channel
calculations on this class of nuclei. To do this, we have deformed the Koning-Delaroche global
spherical potential for neutrons, coupling a sufficient number of states of the ground state band
to ensure convergence. We present an extensive study of the effects of collective couplings and
nuclear deformations on integrated cross sections as well as on angular distributions for
neutron-induced reactions on statically deformed nuclei in the rare-earth region. We choose isotopes
of three rare-earth elements (Gd, Ho, W), which are known to be nearly perfect rotors, to exemplify the
results of the proposed method. Predictions from our model for total, elastic and inelastic cross sections,
as well as for elastic and inelastic angular distributions, are in reasonable agreement with
measured experimental data. These results suggest that the deformed Koning-Delaroche potential
provides a useful regional neutron optical potential for the statically deformed rare earth nuclei.
\end{abstract}
\pacs{24.10.-i, 24.10.Eq, 24.10.Ht, 24.50.+g, 25.40.-h}
\date{\today}

\maketitle

\setstcolor{red}

\section{Introduction}

The optical model has proven over the years to be a powerful method to describe observed nuclear
reaction data \cite{Satchler:1979}. It significantly reduces the complexity of the scattering problem
by employing a complex optical potential that implicitly reproduces the loss of flux due to excitation
of internal degrees of freedom of the nuclei as well as to the opening of inelastic channels.
The parameters of such potentials are often determined by a phenomenological fit to relevant experimental
data either on individual nuclei, nuclei within a region, or globally a large portion of the periodic table.
Over the last several decades, a number of global optical potentials for neutron and proton scattering have
been determined by fitting data on a wide variety of spherical nuclei.  A particularly successful spherical
global potential, which we employ in the present work, was produced by Koning and Delaroche~(KD)~\cite{KD}.

Regions of high static nuclear deformation, such as the ones found in the rare earth and actinide nuclei,
have in general been excluded in the development of global potentials.  Highly deformed nuclei require
a coupled channels~(CC) treatment that accounts for the direct excitation of the rotational states of the
target in order to reproduce experimental data accurately.  It has been conventionally assumed that
potentials used in CC calculations on rotational nuclei must be significantly altered from those for
spherical nuclei, since the inelastic channels treated directly in the CC calculations should no longer
be included implicitly in the optical potential used in these calculations.  On the other hand, recent
work~\cite{Dietrich:2012} has shown that scattering from statically deformed nuclei in the rare-earth and
actinide nuclei is very close to the adiabatic limit.  That is, the nuclei may be regarded as nearly
``frozen" during the scattering process.  This suggests that loss of flux through excitation of
the rotational degrees of freedom might not play a fundamental role in determining the optical potential.
In this paper, we test the hypothesis that a global spherical optical potential, appropriately deformed,
can describe neutron scattering observables in the rare earth region without needing a significant
alteration of its parameters.  An extension of these tests to actinide nuclei will be presented in
a later paper \cite{Dietrich:private}.

The KD neutron global optical potential is particularly suitable for these tests because it has been
successfully fitted to a wide variety of experimental data over a wide energy range (0--200~MeV)
on nuclei with masses both below and above the deformed rare earth region.  Since the KD potential
is parameterized as a smooth function of target mass $A$ and the asymmetry parameter $(N-Z)/A$, we assume
that its interpolation into the rare earths is a useful starting point for the current investigations.
We use the KD potential as the bare potential for coupled-channel calculations without any changes
in its parameterization, except by a small reduction of the radius of the real central part of the potential 
to ensure conservation of its volume integral when it is deformed.

We have carried out calculations of neutron-scattering observables on isotopes of Gd, Ho, and W.
The results indicate that the deformed KD potential produces a fairly satisfactory representation
of the experimental data without further adjustment.  This is particularly true for the real potential,
which determines the angles of the maxima and minima in angular distributions, as well as the positions
of the maxima and minima in the Ramsauer oscillations of the total cross sections.  In many cases, the
back-angle cross sections of the angular distributions are well reproduced by the calculations, whereas
in others they are somewhat underpredicted.  We did not find an easy way to alter the potential
(in particular, its imaginary part) in a manner that varies slowly with mass and also achieves a
fully satisfactory description in all cases.  Nevertheless, the prescription described here appears
to yield a useful regional potential for the deformed rare-earth region that takes advantage of the
extensive physical content already built into the global KD potential.

We note two alternative approaches that have been taken to unify scattering on spherical and
deformed nuclei.  Kunieda~{\it et al.}~\cite{Kunieda:2007} have developed a global phenomenological
potential in which they considered all nuclei as statically deformed, regardless of their actual deformation.
The use of microscopic folding models is promising, since the nuclear densities used in such models may be either spherical or deformed.  In fact, a folding model with an interaction based on the nuclear matter optical potential of Jeukenne, Lejeune, and Mahaux~(JLM)~\cite{Jeukenne:74, Jeukenne:77, Jeukenne:77a}, which is usually carried out in spherical nuclei, has been successfully applied to the deformed rare earth Gd isotopes~\cite{Bauge:2000}.  Similar extensions to deformed nuclei might usefully be carried out with other microscopic treatments, such as those of Refs.~\cite{Nobre:2010,Nobre:2011}.

\section{Coupled-channel model for rare-earths}

The process of deforming a spherical OMP to explicitly consider collective excitations
within the coupled-channel framework is done in the standard way of replacing the radius parameter $R$
in each Woods-Saxon form factor by the angle dependent expression:
\begin{equation}
\label{Eq:DefRadius}
R(\theta)=R_0\left( 1+\sum_\lambda{\beta_\lambda Y_{\lambda0}(\theta)} \right)
\end{equation}
where $R_0$ is the undeformed radius of the nucleus, and $\beta_\lambda$ and $Y_{\lambda0}(\theta)$ are
the deformation parameter and spherical harmonic for the multipole $\lambda$, as described in
Ref.~\cite{Krappe:1976}, for example. The deformed form factor obtained using Eq.~\ref{Eq:DefRadius}
is then expanded in Legendre polynomials numerically.

We use in our calculations the \textsc{Empire} reaction code \cite{Herman:2007,EmpireManual},
in which the direct reaction part is calculated by the code \textsc{Ecis} \cite{Raynal70,Raynal72}.
In previous works \cite{Nobre:2014,Nobre:2014AIP,Herman:2014}, we made preliminary tests of our model
by performing coupled-channel calculations, coupling to the ground state rotational band, for neutron-incident
reactions on selected rare-earth nuclei, namely $^{152,154}$Sm, $^{153}$Eu, $^{155,156,157,158,160}$Gd,
$^{159}$Tb, $^{162,163,164}$Dy, $^{165}$Ho,  $^{166,167,168,170}$Er, $^{169}$Tm,
$^{171,172,173,174,176}$Yb, $^{175,176}$Lu,
$^{177,178,179,180}$Hf, $^{181}$Ta, and $^{182,183,184,186}$W.
All of these nuclides have at least 90 neutrons, which is a reasonable indicator for static deformation,
therefore making them suitable candidates for testing our model based on the approximate validity
of the adiabatic limit. As an initial test, we then compared the coupled-channel results
for total cross sections with plain spherical calculations using the undeformed KD optical potential.
In this initial step, only quadrupole deformations were considered, with values for the deformation
parameters taken from the compilation of experimental values from Raman \emph{et al.} \cite{Raman}.
The overall result, as seen in Refs. \cite{Nobre:2014,Nobre:2014AIP,Herman:2014}, is a very significant
improvement in the agreement with experimental data, in particular for the lower incident-neutron energies
(below about 1~MeV).

\subsection{Radius correction for volume conservation}

When an originally spherical configuration assumes a deformed shape, defined by quadrupole and hexadecupole
deformation parameters $\beta_2$ and $\beta_4$, respectively, the volume and densities are not conserved.
In Ref.~\cite{Bang:1980}, a method
to ensure volume conservation was described, implemented by applying a correction to the nuclear
radius $R_0$, of the form:
\begin{equation}
 R'_{0}=R_0 \Delta_R=R_0\left(  1-\sum^{}_{\lambda}{\beta_{\lambda}^{2}/4\pi}\right) ,
\label{Eq:radius}
\end{equation}
in which $R'_{0}$ is the corrected radius, and where terms of the order of $\beta_{\lambda}^{3}$ and
higher have been discarded. In Ref.~ \cite{Nobre:2014} we tested the effects of such correction, and
showed that while the difference in calculation results is quite small, it is not negligible and seems to bring the integral
and differential cross-section calculations into slightly better agreement with
the experimental data. As a result of these tests, we adopt the radial correction expressed in
Eq.~\ref{Eq:radius} in the following calculations and further tests of the model.

\subsection{Compound-nucleus parameterization}

Even though the model being tested in this work probes the direct-reaction mechanism,
it is also important to obtain a reliable description of the processes involved after the formation
of the compound nucleus, following neutron absorption. Such compound contributions to the elastic and
inelastic channels are much smaller than their direct-reaction correspondents
(shape elastic and direct inelastic excitation) at sufficiently high energies, but make significant
contributions to some of the reactions we consider here at low incident energies
(typically in the neighborhood of 1--2~MeV or less).

%After the initial success in describing direct-reaction quantities, such as total cross sections, we analyzed the model predictions for observables that depend also on the compound-nucleus decay.
The models adopted to describe the emissions from the compound nucleus were basically
the standard options within the \textsc{Empire} code, which means standard
Hauser-Feshbach model with properly parameterized Enhanced Generalized Superfuid Model (EGSM)
level densities \cite{fade,DArrigo:1992}, modified Lorentzian distribution (version 1)
for $\gamma$-ray strength functions \cite{plu01,plu02,plu03}, width fluctuation correction
implemented up to 3 MeV in terms of the HRTW approach \cite{HRTW,HHM}, and with transmission
coefficients for the inelastic outgoing channels also calculated within the coupled-channel
approach (the KD potential was also used in outgoing channels). Pre-equilibrium was calculated
within the exciton model~\cite{Griffin:66}, as based on the solution of the
master equation~\cite{Cline:71} in the form proposed by Cline~\cite{Cline:72}
and Ribansk\'{y}~\cite{Ribansky:73} (using the \textsc{Pcross} code \cite{Herman:2007,EmpireManual})
with mean free path multiplier set to 1.5.

\section{Tests of the proposed model for scattering from statically deformed rare earth nuclei}

In this work we present the results of integral and differential cross sections
for $^{158,160}$Gd, $^{165}$Ho, and $^{182,184,186}$W obtained by our model, following the
preliminary results shown in Refs.~\cite{Nobre:2014AIP,Herman:2014}. The reason for choosing
the three elements (Gd, Ho, W) for the testing of our model is that they approximately span
the statically deformed part of the rare-earth region, and there are suitable experimental
data available for comparison with calculations.  The values for deformation parameters that
were adopted for the different isotopes are shown in Table~\ref{Tab:Deformations}, as well as
the radius correction used, calculated from Eq.~(\ref{Eq:radius}).  In addition to the $\beta_2$
values extracted from analyses of neutron scattering experiments, we also show the values from the
compilation of Raman {\it et al.}~\cite{Raman} based on electromagnetic B(E2) data for even-even nuclei.
The values obtained from scattering experiments are systematically smaller than those from
the Raman {\it et al.} compilation, and we have found that they are better in reproducing
the magnitudes of inelastic excitations.  However, in cases where an independent determination
of $\beta_2$ from scattering is not available, the values Raman compilation may be useful if
appropriately scaled. Systematics, such as the ones from Refs.~\cite{Chamon:2004,Nobre:2007}, may be useful to retrieve $\beta_\lambda$ values in cases where experimental data are not available. In some cases, particularly at low energies ($\lesssim 1$~MeV), the accuracy of the calculations is strongly dependent on the accuracy of the adopted deformation parameters.

\begin{table}[h]
\caption{Deformation parameters and radial corrections used in coupled-channel calculations. References
from which the values were taken are also indicated below.}
\label{Tab:Deformations}
\begin{ruledtabular}
\begin{tabular}{lcccr}
 Nuclide & $\Delta_R$ & $\beta_2$ & $\beta_4$ & $\beta_2^{\mathrm{Raman}}$  \\
\hline \vspace{-2mm} \\
$^{155}$Gd & 0.995 & 0.25 \cite{Bauge:2000}                                & +0.07 \cite{Bauge:2000}                                &                          \\
$^{156}$Gd & 0.995 & 0.25 \cite{Bauge:2000}                                & +0.06 \cite{Bauge:2000}                                & 0.3378(18) \cite{Raman}  \\
$^{157}$Gd & 0.994 & 0.26 \cite{Bauge:2000}                                & +0.05 \cite{Bauge:2000}                                &                          \\
$^{158}$Gd & 0.994 & 0.27 \cite{Bauge:2000}                                & +0.04 \cite{Bauge:2000}                                & 0.3484(17) \cite{Raman}  \\
$^{159}$Gd & 0.994 & 0.28 \cite{Bauge:2000}                                & +0.03 \cite{Bauge:2000}                                &                          \\
$^{160}$Gd & 0.993 & 0.29\footnotemark[1]                                      & +0.02\footnotemark[1]                                       & 0.353 \cite{Raman}       \\
$^{165}$Ho & 0.993 & 0.300 \cite{Smith:2001}                               & -0.020 \cite{Smith:2001}                               &                          \\
$^{182}$W  & 0.996 & 0.223 \cite{Guenther:1982,Delaroche:1981,Annand:1985} & -0.055 \cite{Annand:1985}                              & 0.2508(24) \cite{Raman}  \\
$^{184}$W  & 0.996 & 0.209 \cite{Guenther:1982,Delaroche:1981,Annand:1985} & -0.056 \cite{Guenther:1982,Delaroche:1981,Annand:1985} & 0.2362(41) \cite{Raman}  \\
$^{186}$W  & 0.996 & 0.203 \cite{Guenther:1982,Delaroche:1981}             & -0.057 \cite{Guenther:1982,Delaroche:1981}             & 0.2257(39) \cite{Raman}  \\
%$^{186}$W      & 0.995 &       & -0.080 \cite{Hendrie:1968} & 0.226 \cite{Raman}  \\
\end{tabular}
\end{ruledtabular}
\footnotetext[1]{Values obtained by linearly extrapolating the deformation parameters of the lighter Gd isotopes.}
\end{table}

\subsection{Integral cross sections}

\begin{figure}[h]
\begin{center}
\subfigure{\label{Fig:Ho165total}\includegraphics[scale=.5,clip, trim= 5mm 16mm 5mm 10mm]{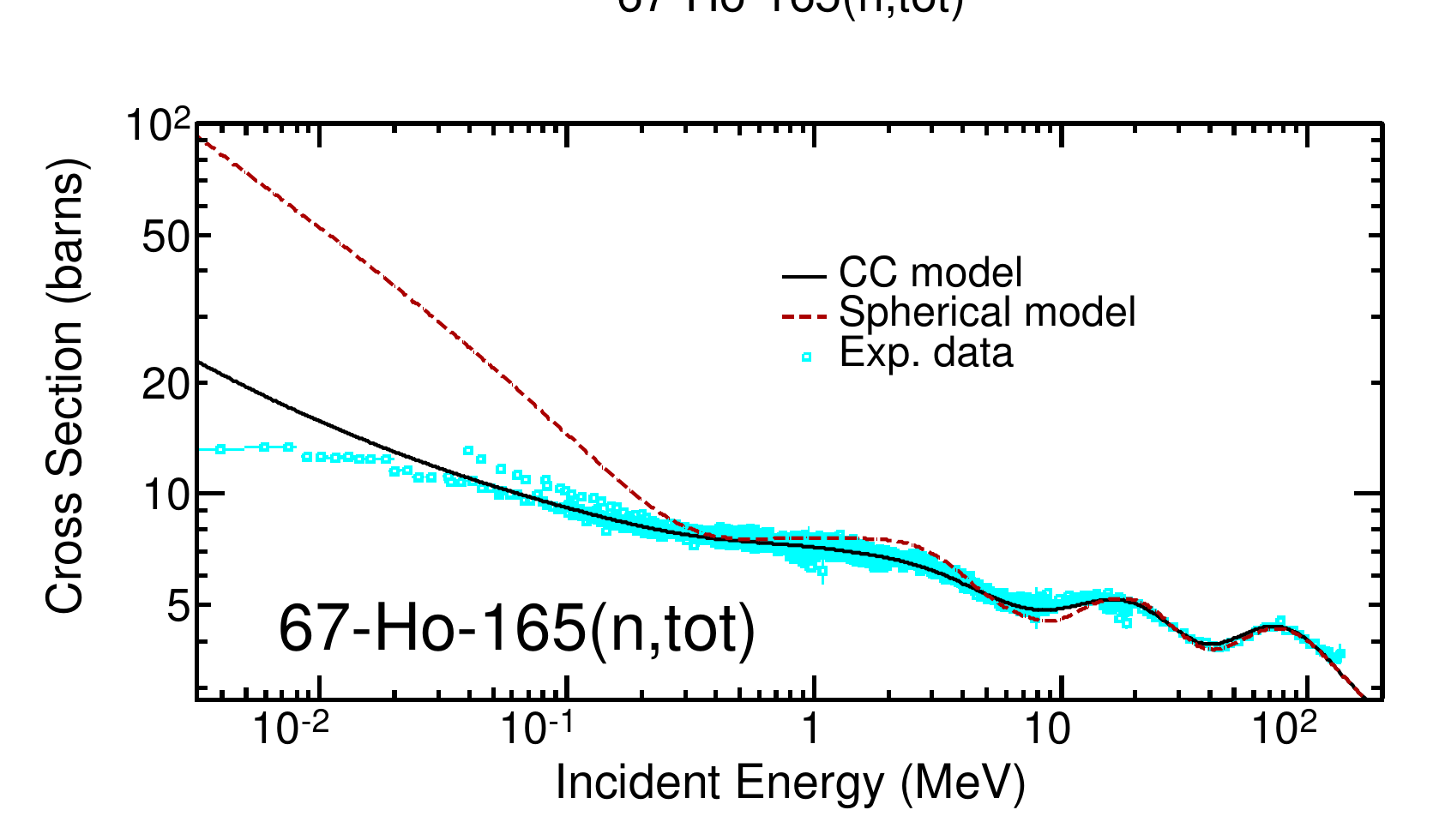}} \\ \vspace{-6.4mm}
\subfigure{\label{Fig:W182total}\includegraphics[scale=.5,clip, trim= 5mm 16mm 5mm 10mm]{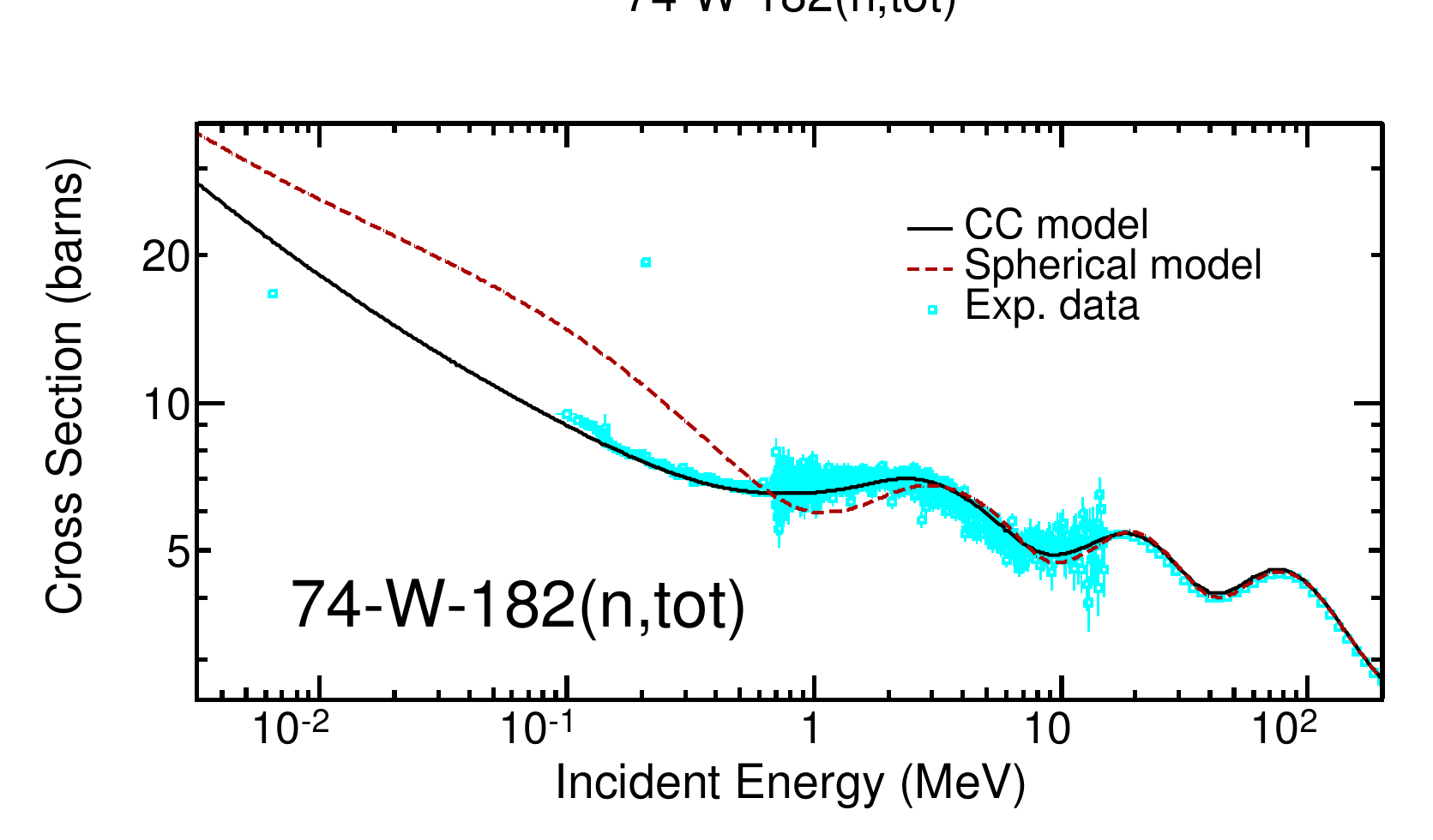}}\\ \vspace{-6.4mm}
\subfigure{\label{Fig:W184total}\includegraphics[scale=.5,clip, trim= 5mm 16mm 5mm 10mm]{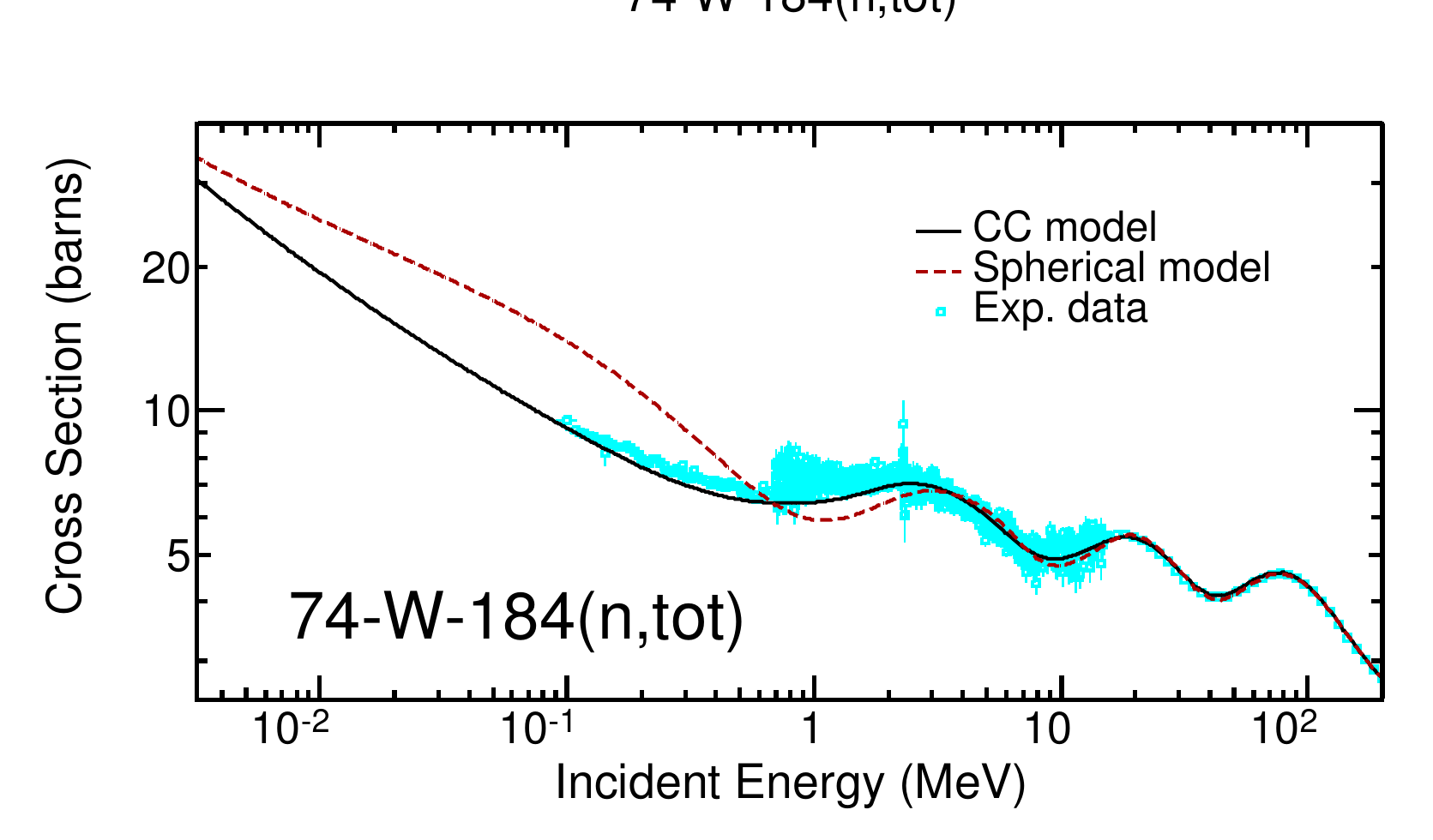}}\\ \vspace{-6.4mm}
\subfigure{\label{Fig:W186total}\includegraphics[scale=.5,clip, trim= 5mm 3mm 5mm 10mm]{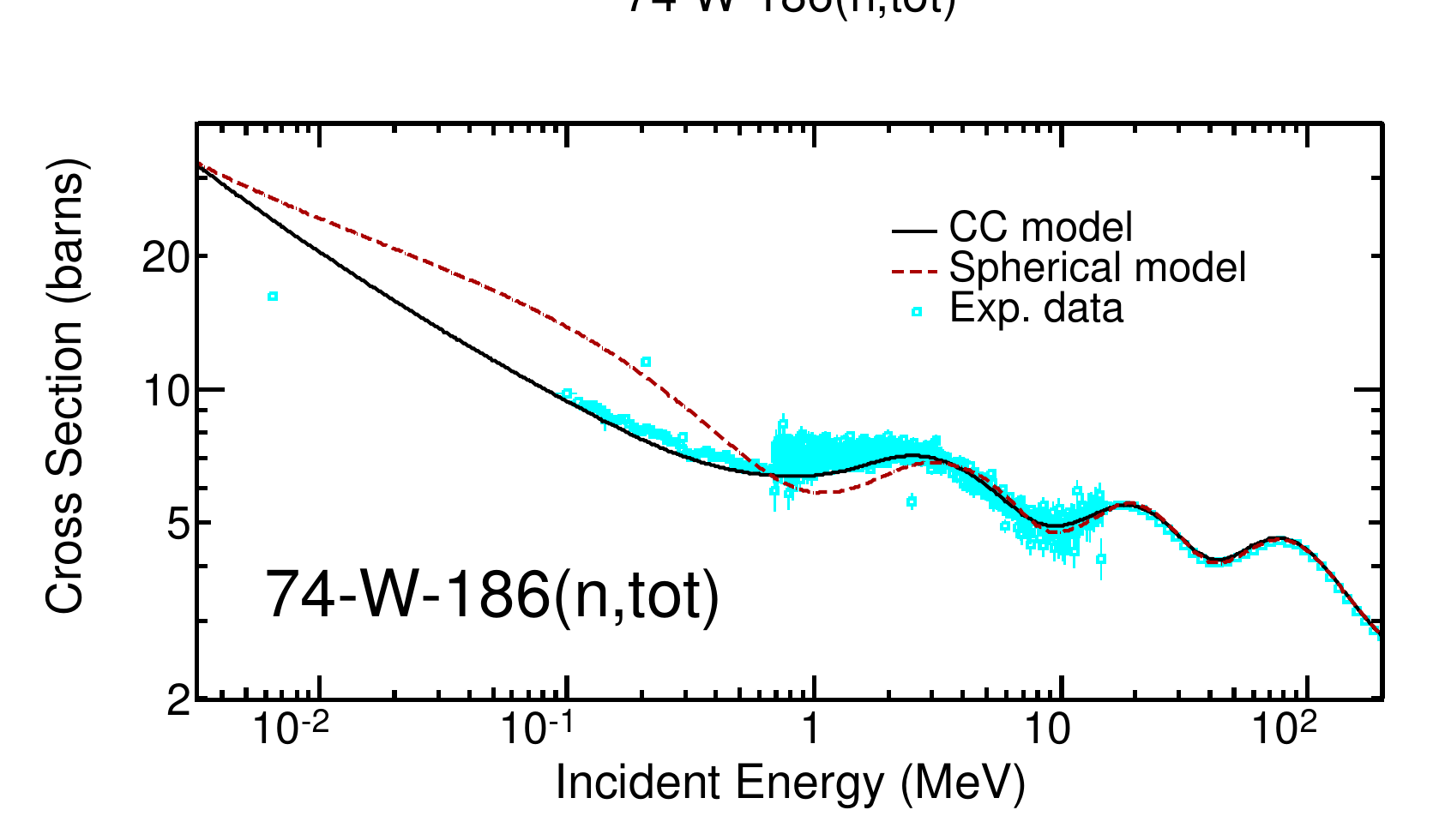}}
\end{center}
\caption{(color online) Total cross sections for neutrons scattered by
a $^{165}$Ho and $^{182, 184, 186}$W targets,
for incident energies ranging from as low as $\approx$3~keV to as high as 200~MeV, which is
the upper limit of validity
for the Koning-Delaroche optical potential \cite{KD}. The solid black curves correspond to the predictions of our
couple-channel model while the dashed red curves are the result of calculations within the spherical model.
The experimental data were taken from the EXFOR nuclear data library \cite{EXFOR}.
%, with the particular data sets indicated in the figure legends.
}
\label{Fig:Total}
\end{figure}

As a straightforward test of our model we calculated the total cross sections for the
reaction of neutrons scattered by the nuclei presented in Table \ref{Tab:Deformations}.
To illustrate our results we present in Fig.~\ref{Fig:Total} the total cross sections for
$^{165}$Ho and $^{182, 184, 186}$W. We can immediately see that, while the spherical model
poorly describes the measured shape of total cross sections, particularly overestimating the
lower-energy region (for some
rare-earth nuclei this difference can be of almost an order of magnitude \cite{Nobre:2014}),
our coupled-channel model based on the approximate validity of the adiabatic
limit yields very good agreement, from lower- to higher-energy regions.
Integral cross sections are presented only up to  200 MeV.
%One noteworthy aspect rises when we notice, in the tungsten cases, that both calculations begin to fail at $E_{\mathrm{inc}}\gtrapprox 200$ MeV.
This is due to the fact that the global spherical Koning-Delaroche potential was fitted to
data at incident energies below 200~MeV, thus only being reliable in this region.

%\begin{figure}
%\begin{center}
 %\includegraphics[height=.28\textheight,,clip, trim= 5mm 3mm 5mm 10mm]{Ho165-total}
%\end{center}
%\caption{(color online) Total cross sections for neutrons scattered by a $^{165}$Ho target, for incident energies ranging from $\approx$30~keV up to 140~MeV. The solid black curve corresponds to the predictions of our couple-channel model while the dashed blue curve is the result of calculations within the spherical model. The experimental data were taken from the EXFOR nuclear data library \cite{EXFOR}, with the particular data sets indicated in the figure legend.}
%\label{Fig:Ho165Total}
%\end{figure}

\subsection{Angular distributions}

To better assess the quality and effectiveness of  our coupled-channel model we compared
its predictions to a variety of experimental angular distribution data. Such differential
data are typically more sensitive to details of the optical potential and the deformations than the integral cross sections.

\subsubsection{Gadolinium isotopes}

We compare our calculations with two sets of angular distributions, those measured by Bauge \emph{et al.}
\cite{Bauge:2000} and by Smith \emph{et al.} \cite{Smith:2004}. In the former case, it was possible
to separate the differential cross sections of the elastic and first two inelastic channels for the
two heavier stable isotopes, for two different incident energies; while in the latter one measurements
were presented at several incident energies for natural Gadolinium, without being able to resolve
the inelastic contributions in most cases.

\begin{figure*}
\begin{center}
   \subfigure[$^{158}$Gd angular distributions.]{\label{Fig:Gd158Bauge} \includegraphics[height=.5\textheight,,clip, trim= 38mm 22mm 34mm 19mm]{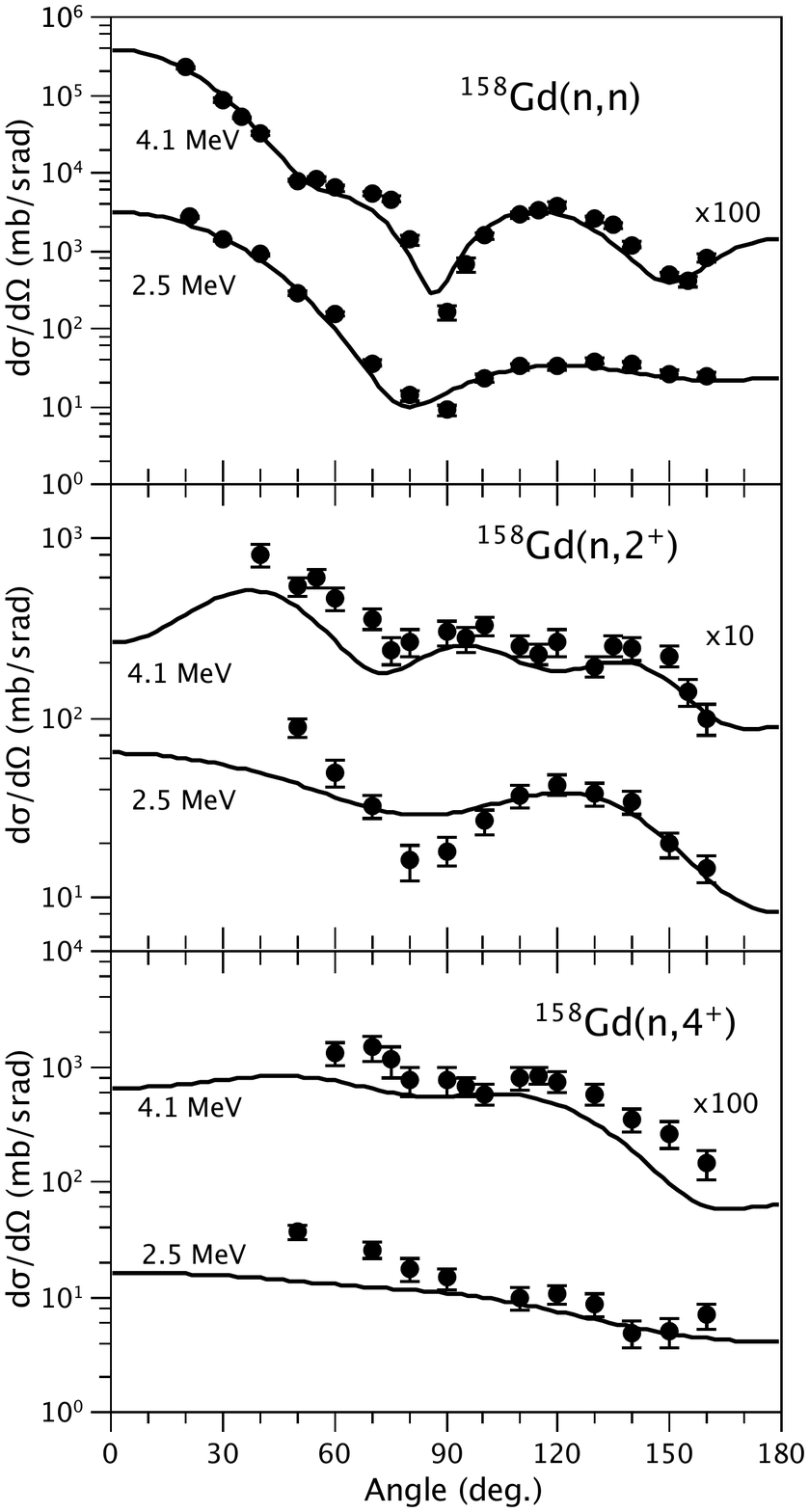}}
   \subfigure[$^{160}$Gd angular distributions.]{\label{Fig:Gd160Bauge} \includegraphics[height=.5\textheight,,clip, trim= 53mm 22mm 34mm 19mm]{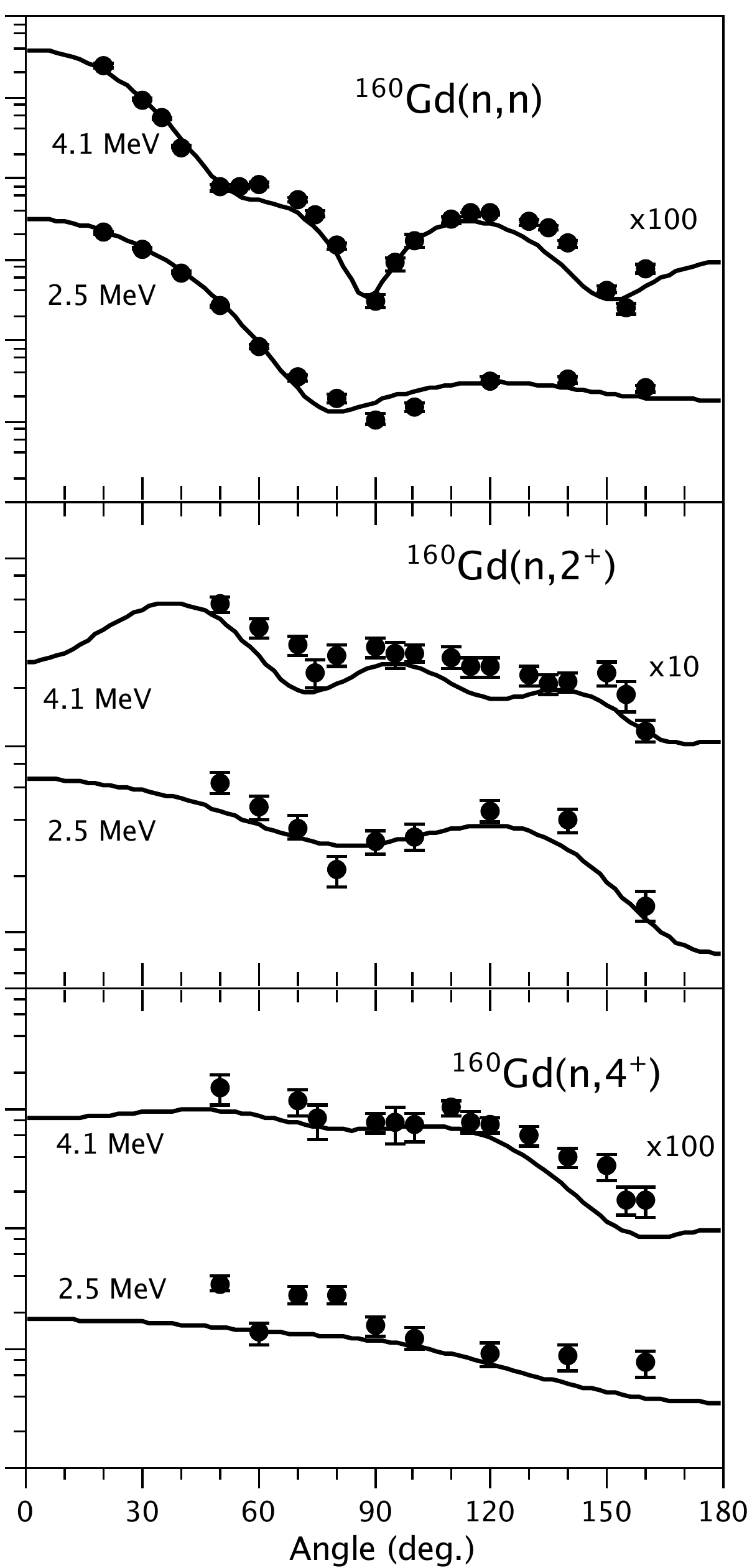}}
\end{center}
\caption{Angular distributions for elastic (upper panels) and first and second inelastic
(middle and bottom panels, respectively)  channels for the neutron-induced reaction on $^{158}$Gd
and $^{160}$Gd. The curves correspond to predictions by our coupled-channel model. Numbers on the
left of each plot indicate, in MeV, the values of incident energy at which the cross sections were
measured, while the numbers on the right side correspond to the multiplicative factor applied to facilitate plotting data from different incident energies in the same graph. Experimental data taken from
Bauge \emph{et al.} \cite{Bauge:2000}.}
\label{Fig:Gd158Gd160Bauge}
\end{figure*}

In Fig.~\ref{Fig:Gd158Gd160Bauge} we compare the predictions of our model for the elastic and
inelastic angular distributions for $^{158}$Gd (Fig.~ \ref{Fig:Gd158Bauge}) and $^{160}$Gd
(Fig.~ \ref{Fig:Gd160Bauge}) with the experimental data measured by Bauge \emph{et al.} \cite{Bauge:2000}.
The elastic differential cross sections are presented in the top panels, while angular
distributions for the first 2$^+$ and 4$^+$ states are shown in the middle  and bottom panels,
respectively. These states have excitation energies ($E^*$) of $E^{*}_{2^+}$=79.5~keV
and $E^{*}_{4^+}$=261.5~keV for $^{158}$Gd and $E^{*}_{2^+}=75.3$~keV and $E^{*}_{4^+}=248.5$~keV
in the case of  $^{158}$Gd.
It can be clearly seen in Fig.~\ref{Fig:Gd158Gd160Bauge} that our model succeeds in reproducing
very well the observed elastic differential cross section (upper panel) for both Gadolinium isotopes studied.
 Regarding the predictions of our coupled-channel model for the angular distributions of the
 first two excited states, as shown in Fig.~\ref{Fig:Gd158Gd160Bauge} (middle and bottom panels),
 even though the agreement with experimental data is not as good as in the case of elastic scattering,
 they still describe reasonably well the measured data, in particular their shape.

Smith \emph{et al.} provides, in Ref.~\cite{Smith:2004}, angular distribution data for natural
(or elemental) Gadolinium ($^{\mathrm{Nat}}$Gd) for neutron incident energies ($E_{\mathrm{inc}}$)
ranging from 0.334 MeV up to 9.99 MeV. For the lower values of incident
energy ($E_{\mathrm{inc}}\lessapprox$ 1 MeV) the elastic channel is completely resolved while the
data sets with higher incident energies (4.51 MeV $\leqslant E_{\mathrm{inc}} \leqslant$ 9.99 MeV)
do not have any separation between elastic and inelastic contributions from members of the ground-state band. Datasets with incident energies
around 1 MeV (1.080 MeV $\leqslant E_{\mathrm{inc}} \leqslant$ 1.432 MeV), did have inelastic
contributions but it was not clear from Ref.~\cite{Smith:2004} which channels were not resolved from
the elastic one. For this reason, we decided not to present comparisons with these data sets. To obtain
theoretical predictions for $^{\mathrm{Nat}}$Gd we proportionally combined the results of calculations
for $^{155}$Gd (14.80\%), $^{156}$Gd (20.47\%), $^{157}
$Gd (15.65\%), $^{158}$Gd (24.84\%), and $^{160}$Gd (21.86\%), according to their contribution to the
natural occurence of the element, as indicated in parentheses.
$^{152}$Gd (0.20\%) and $^{154}$Gd (2.18\%)
were ignored due to their small contribution (less than 3\%) and normalization was done accordingly.

\begin{figure}
\begin{center}
 \includegraphics[height=.28\textheight,,clip, trim= 8mm 71mm 5mm 68mm]{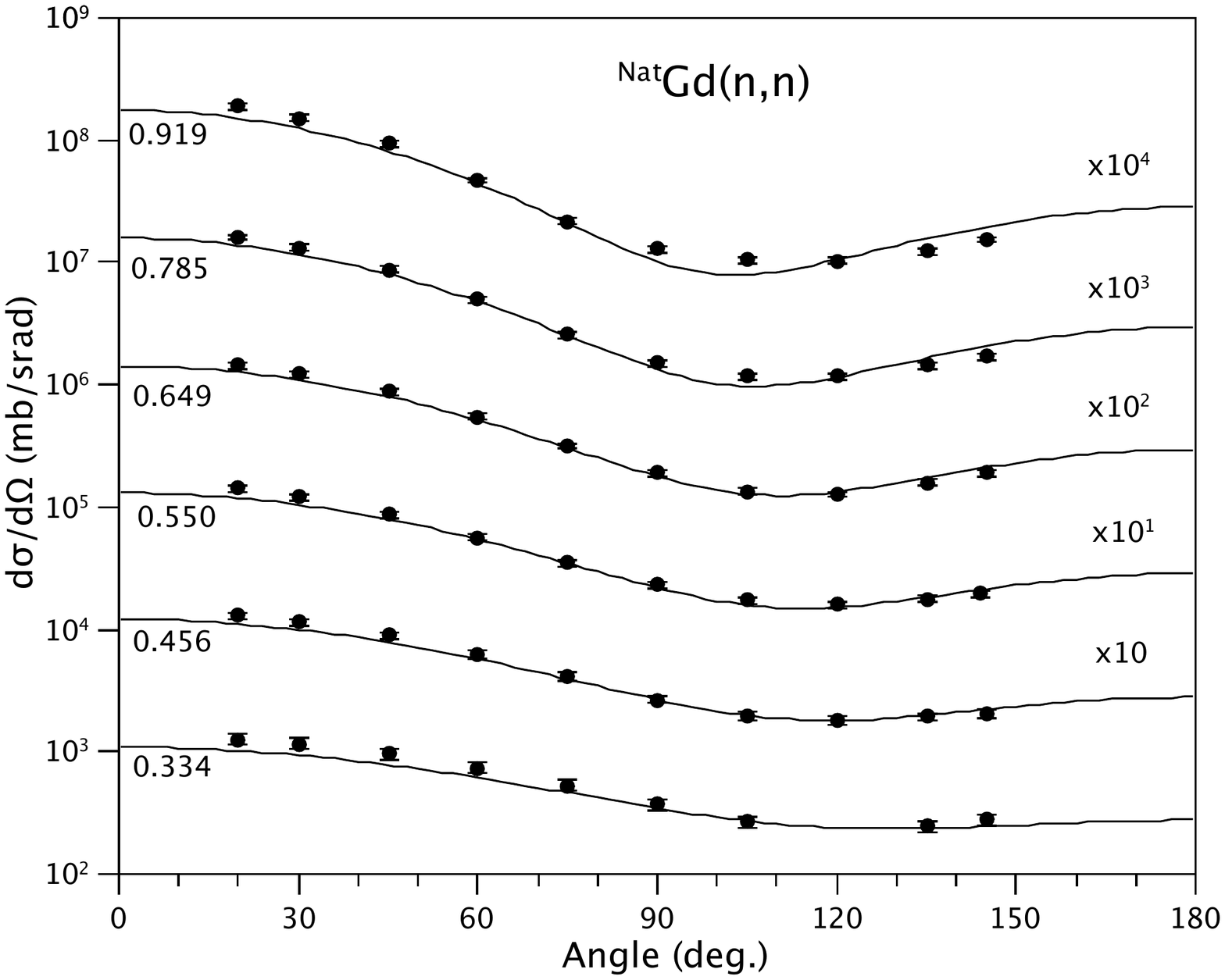}
\end{center}
\caption{Elastic angular distributions for neutron-induced reactions on $^{\mathrm{Nat}}$Gd.
The curves correspond to predictions by our coupled-channel model. Numbers on the left of each plot
indicate, in MeV, the values of incident energy at which the cross sections were measured, while the numbers
on the right side correspond to the multiplicative factor applied to be able to plot data from different
incident energies in the same graph. Experimental data taken from Smith \emph{et al.} \cite{Smith:2004}.}
\label{Fig:GdNatSmithLow}
\end{figure}

Fig.~\ref{Fig:GdNatSmithLow} shows the predictions of our model for the elastic angular
distribution of $^{\mathrm{Nat}}$Gd which are in excellent agreement with the observed data.
Very small discrepancies are more apparent only for the highest incident energy ($E_{\mathrm{inc}}$=0.919),
probably because those data are beginning to incorporate some inelastic contributions.

\begin{figure}
\begin{center}
 \includegraphics[height=.5\textheight,,clip, trim= 11mm 12mm 9mm 7mm]{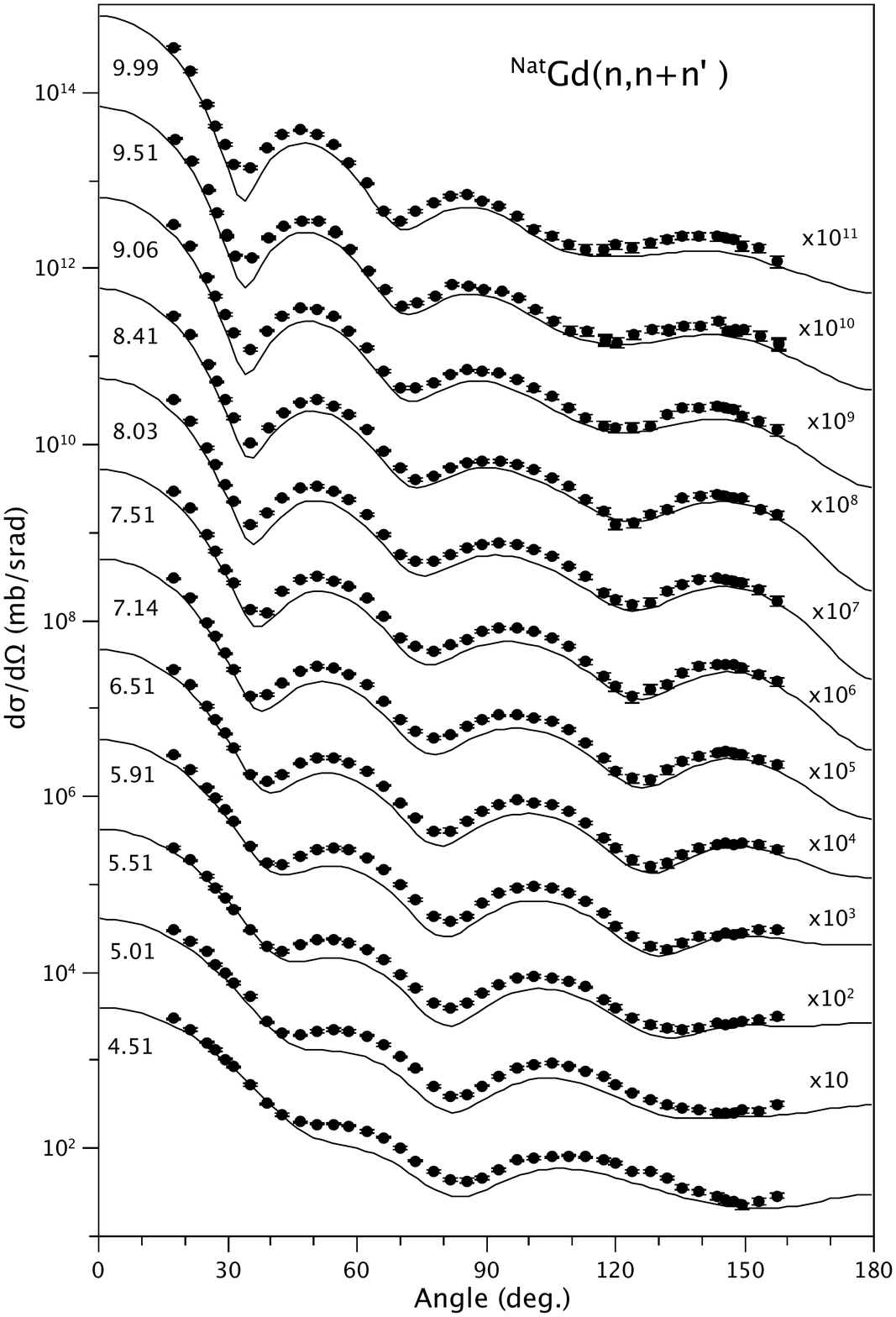}
\end{center}
\caption{Sum of elastic and inelastic differential cross sections for neutrons
scattered by $^{\mathrm{Nat}}$Gd. The curves correspond to predictions by our coupled-channel model.
Numbers on the left of each plot indicate, in MeV, the values of incident energy at which the
cross sections were measured, while the numbers on the right side correspond to a multiplicative offset.
Experimental data taken from Bauge \emph{et al.} \cite{Bauge:2000}.}
\label{Fig:GdNatSmithHigh}
\end{figure}

We present in Fig.~\ref{Fig:GdNatSmithHigh} the results of our model for the summed differential
cross sections of the elastic channel with the contributions from the first four excited states.
That means the first 2$^+$, 4$^+$, 6$^+$ and 8$^+$ in the case of $^{156}$Gd, $^{158}$Gd and $^{160}$Gd.
For such even-even nuclei the most significant inelastic contributions come from the 2$^+$ and  4$^+$ while
the higher members of the ground state (g.\ s.) band make successively smaller contributions.
For $^{155}$Gd (which has a 1/2$^-$ g.\ s.\ state) we added the contributions from the
first 3/2$^-$, 5/2$^-$, 7/2$^-$ and 9/2$^-$ states, while in the case of $^{157}$Gd (3/2$^-$ g.\ s.)
we summed up the differential cross sections from the  5/2$^-$, 7/2$^-$, 9/2$^-$ and 11/2$^-$
inelastic channels. For clarification purposes, it is important to state that, even though we considered
only the contributions from the first four inelastic channels in Fig.~\ref{Fig:GdNatSmithHigh},
we obviously performed calculations
coupling to a much larger number of inelastic states of the g.\ s.\ band in order to absolutely
ensure convergence regarding the number of channels coupled. The comparisons presented in
Fig.~\ref{Fig:GdNatSmithHigh} show that we successfully predict the observed shape of such summed
%``quasi-elastic''
differential cross sections, even though our calculations tend to
slightly underestimate the angular distributions, especially for lower incident energies. Since we
can assume, from Figs.~\ref{Fig:Gd158Gd160Bauge} and~\ref{Fig:GdNatSmithLow} that we obtain an excellent
description of the elastic channel, this small discrepancy may be due to the need of fine tuning of
the calculation of the inelastic channels associated with quadrupole and hexadecupole deformations.  We did not find that altering the strength of the imaginary potential within reasonable limits ($\pm$15\%) significantly improved the overall agreement.  Given that there has been no adjustment of the KD optical parameters (other that the radius change to impose volume conservation), nor of the deformations, the results are satisfactory.

\subsubsection{Holmium}

The ground state of $^{165}$Ho has spin and parity 7/2$^-$, and consequently a ground-state band level sequence 7/2$^-$, 9/2$^-$, 11/2$^-$, 13/2$^-$, $\cdots$~.  Coupled-channel calculations were
performed coupling all states of the g.s. band up to 23/2$^-$ to ensure convergence. Due to the experimental difficulties of resolving the
elastic channel from the inelastic ones in neutron-induced reactions on $^{165}$Ho,
experimental data sets for higher incident energies (above $\approx$~1~MeV)
usually contain inelastic contributions.

\begin{figure}
 \begin{center}
  \includegraphics[height=.37\textwidth,clip, trim= 7mm 73mm 0mm 71mm]{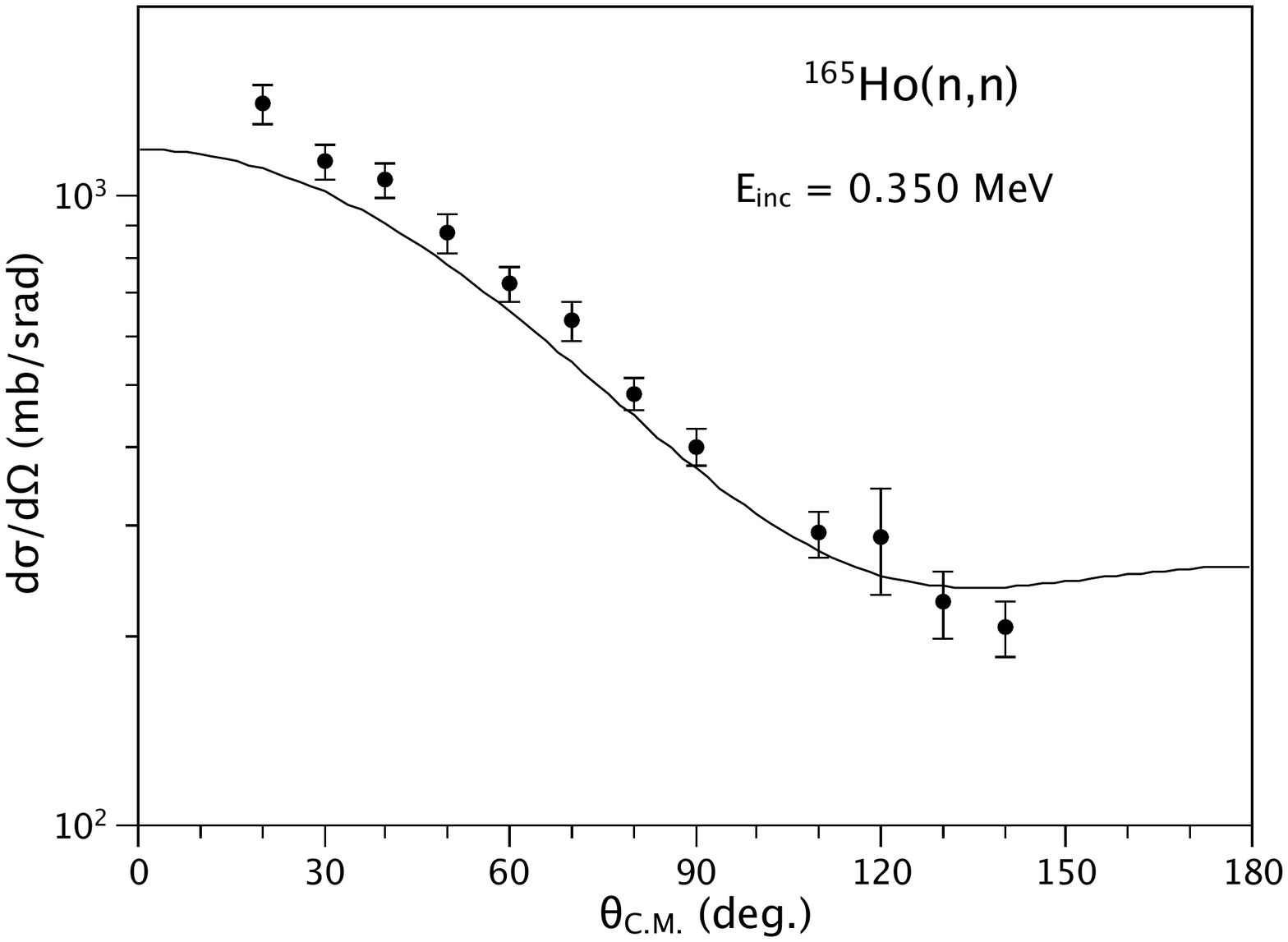}
 \end{center}
  \caption{Elastic angular distribution for the neutron-induced reaction on $^{165}$Ho
  at the incident energy of 0.350 MeV. The curve corresponds to the elastic-channel
  results obtained within our coupled-channel model. Experimental data taken from
  Ref.~\cite{Wagner:1965}.}
 \label{Fig:Ho165-Wagner}
\end{figure}

Fig.~\ref{Fig:Ho165-Wagner} shows the experimental data measured by Wagner \emph{et al.} \cite{Wagner:1965}
for the incident energy of 0.350 MeV, which correspond to pure elastic differential
cross sections. As can be seen in Figure~\ref{Fig:Ho165-Wagner}, the predictions of
our model for the elastic channel (black curve)  %are enough to
describe the observed
angular distributions very well.

\begin{figure}
\begin{center}
   \subfigure{\includegraphics[scale=0.22,,clip, trim= 6mm 104mm 3mm 69mm]{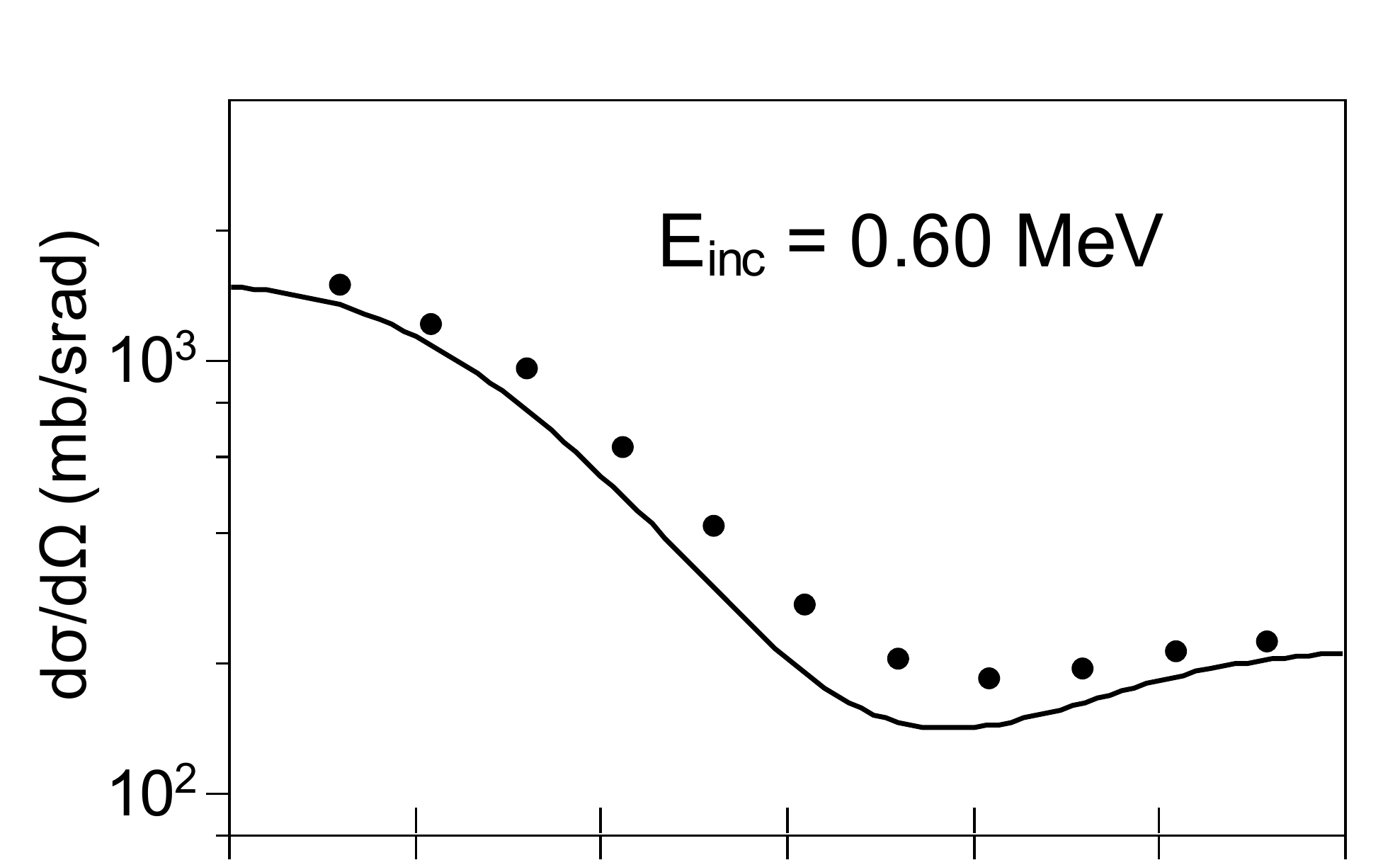}} \hspace{-4.1mm}
   \subfigure{\includegraphics[scale=0.22,,clip, trim= 9mm 104mm 21mm 69mm]{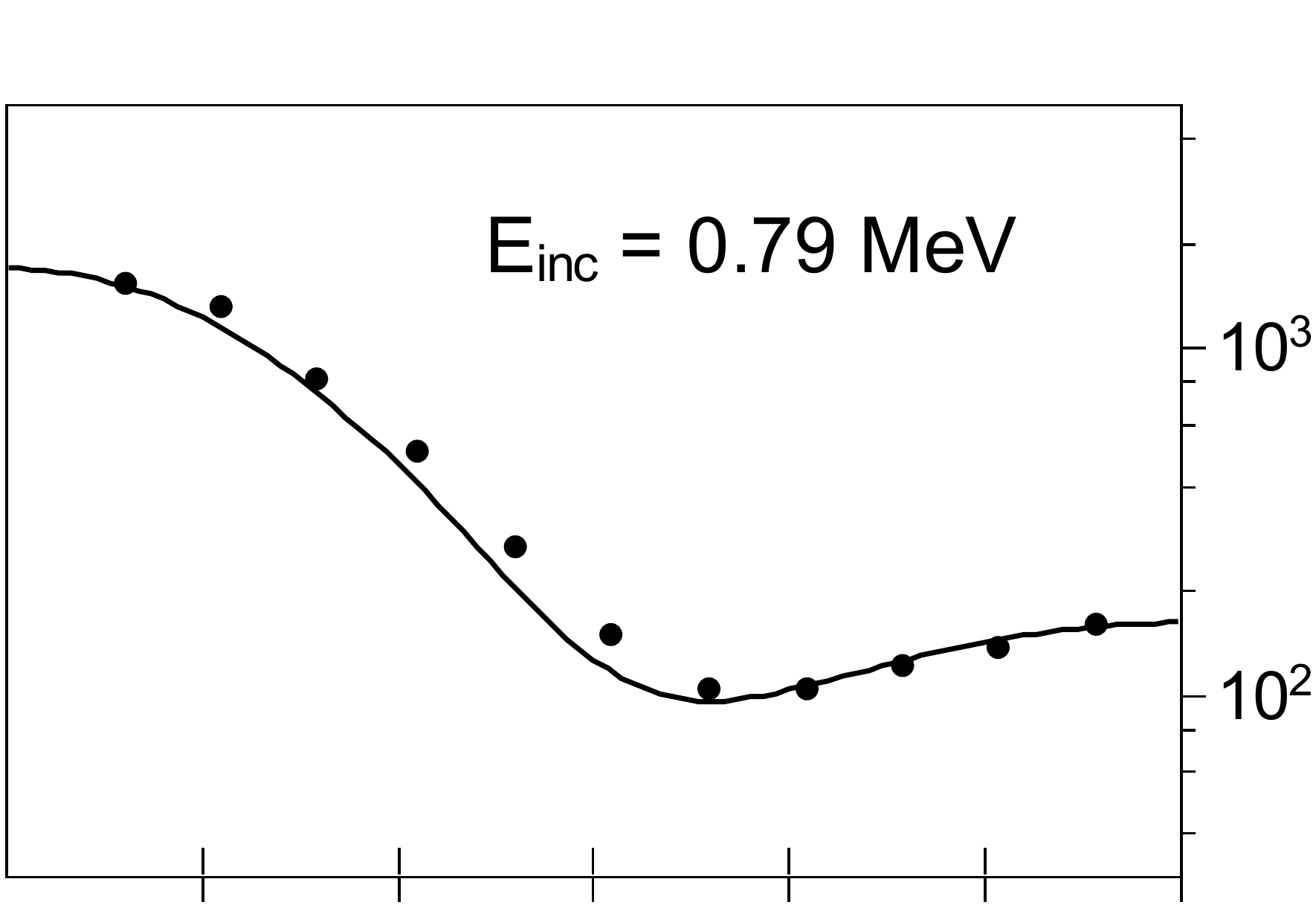}}\\ \vspace{-5.8mm}
   \subfigure{\includegraphics[scale=0.22,,clip, trim= 6mm 72mm 3mm 79mm]{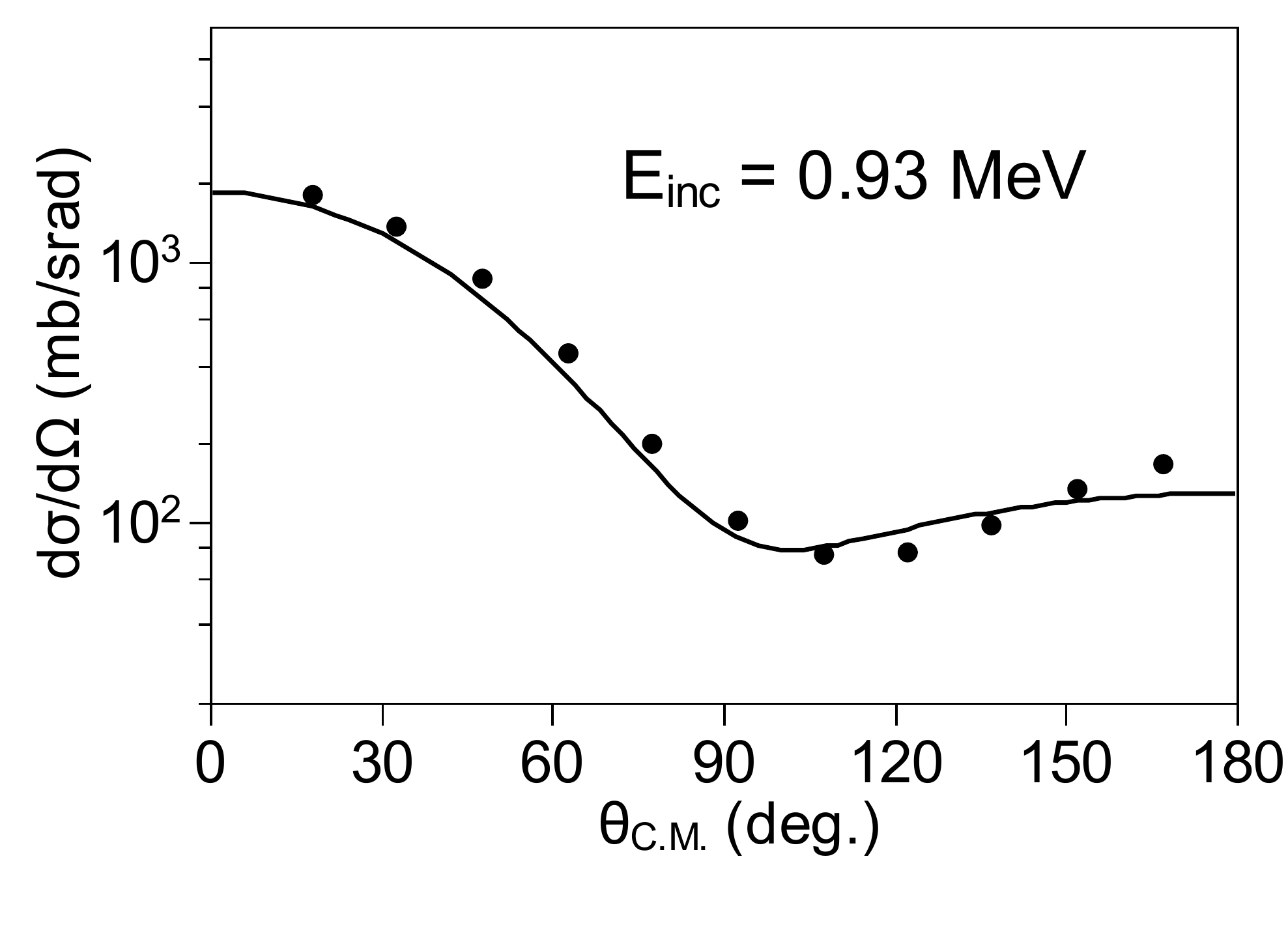}} \hspace{-4.1mm}
   \subfigure{\includegraphics[scale=0.22,,clip, trim= 9mm 72mm 21mm 79mm]{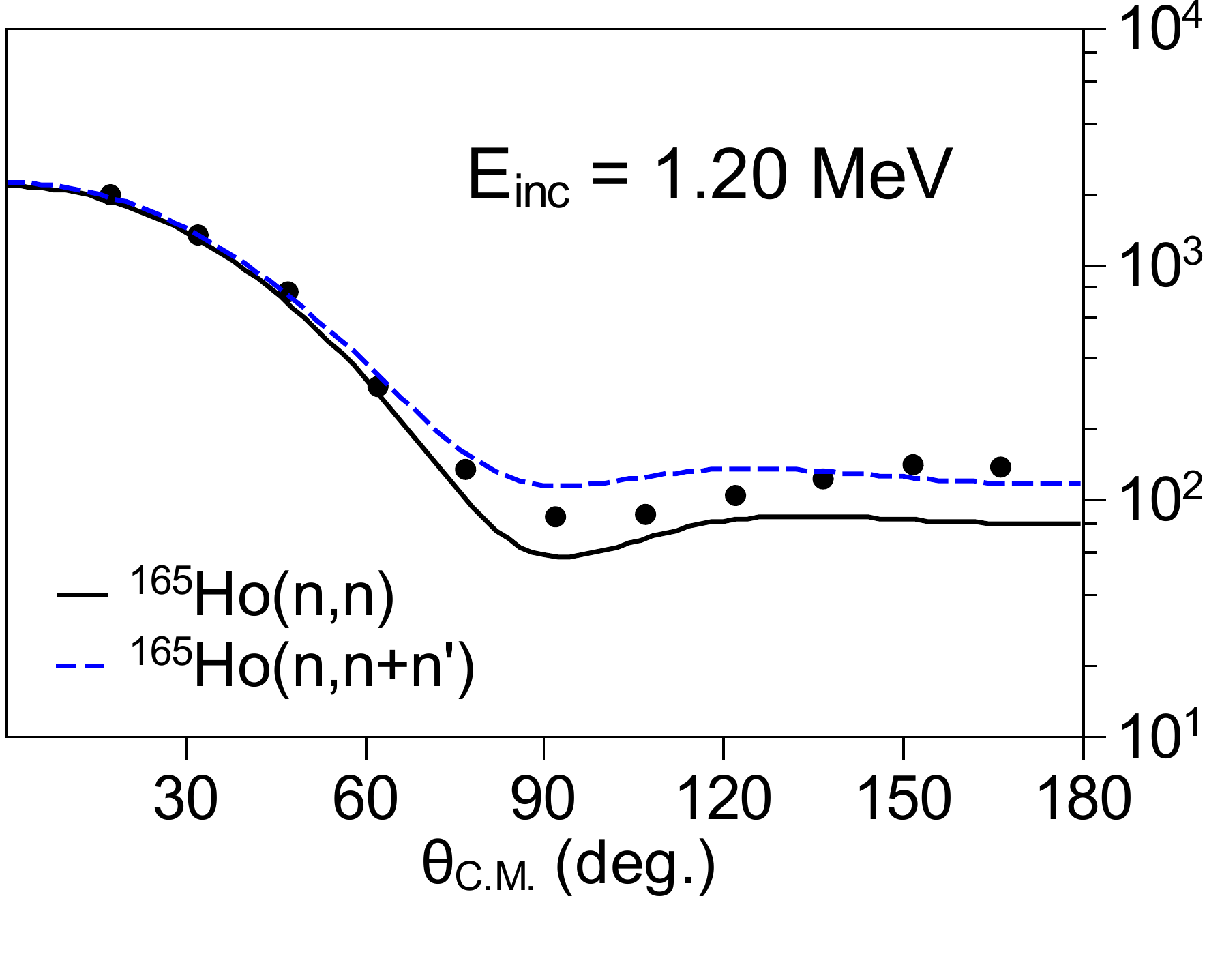}}
\end{center}
\caption{(color online) Comparison of calculations using our coupled-channel
model with measured angular distributions for the neutron-induced reaction on $^{165}$Ho.
The black curves correspond to the elastic-channel results while the blue dashed curve
corresponds to the elastic differential cross sections with the added contributions from
inelastic channels. Even though the measurements were described in Ref.~\cite{Meadows:1971},
the  experimental data were actually taken directly from Ref.~\cite[Fig.~3]{Smith:2001}.}
\label{Fig:Ho165Meadows}
\end{figure}

We also assessed %the quality of
the agreement of elastic and elastic plus inelastic angular distributions
calculated from our model with early experimental measurements from
Meadows \emph{et al.} \cite{Meadows:1971}, done for incident energies between $\approx$ %around
0.3 to 1.5 MeV. Ref.~\cite{Meadows:1971} presents its results in the form of Legendre
expansions of the angular distributions, while Ref.~\cite{Smith:2001} reconstructs the
corresponding differential cross sections. In Fig.~\ref{Fig:Ho165Meadows} we present such
comparisons for selected values of incident energy, namely, $E_\mathrm{inc}=0.60$, 0.79, 0.93
and 1.20 MeV. Again, as can be seen in Fig.~\ref{Fig:Ho165Meadows}, we obtain a very good
description of the experimental data with our coupled-channel model. In the case
of $E_\mathrm{inc}=1.20$ MeV (lower right-hand panel of Fig.~\ref{Fig:Ho165Meadows}),
for which the experimental data contain inelastic contributions \cite{Smith:2001}, we also show
the resulting calculation of summing up elastic and inelastic angular
distributions.

\begin{figure}
\begin{center}
   \subfigure{\includegraphics[scale=0.45,,clip, trim= 9mm 104mm 3mm 69mm]{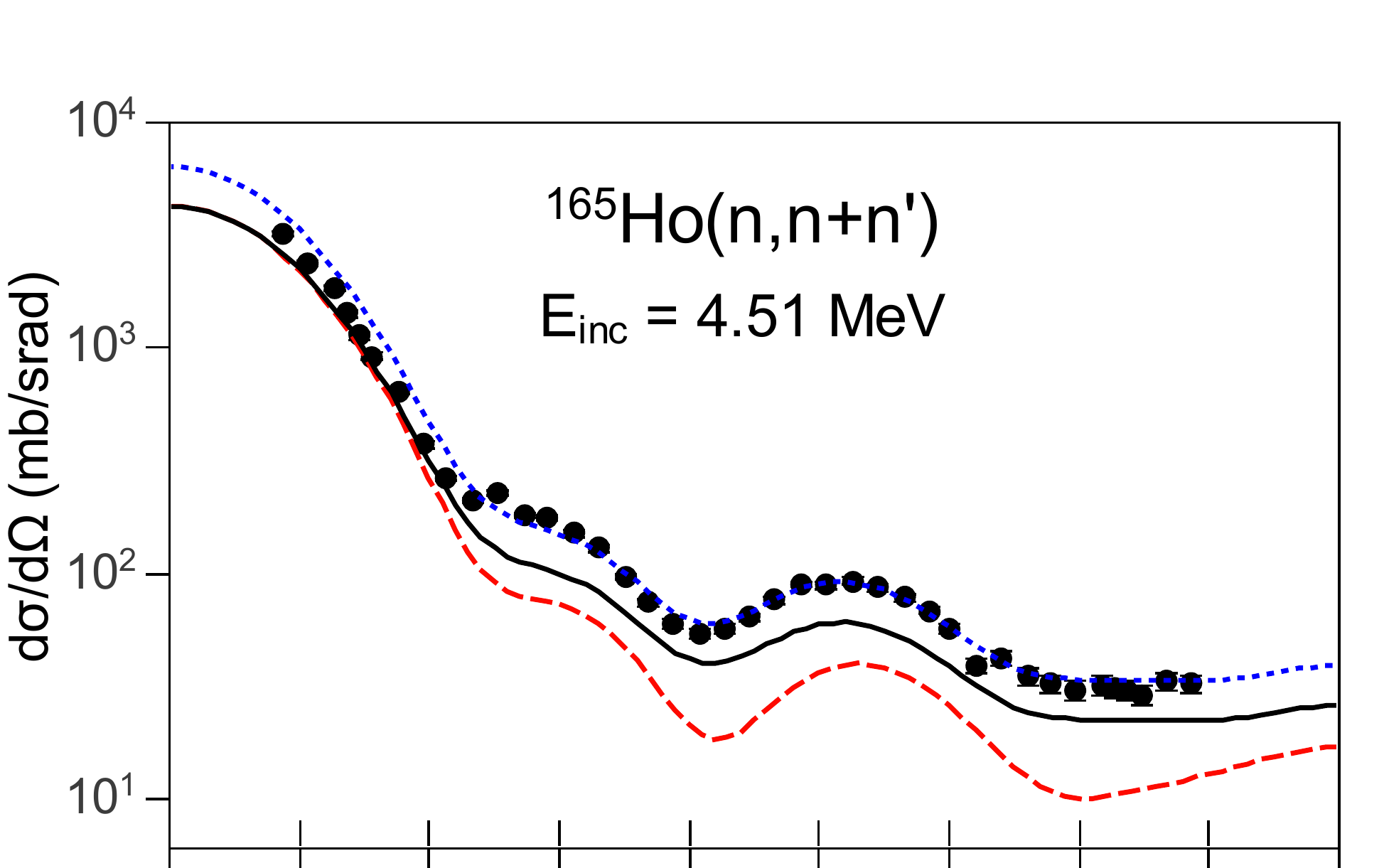}}\\ \vspace{-8.5mm}
   \subfigure{\includegraphics[scale=0.45,,clip, trim= 9mm 86mm 3mm 79mm]{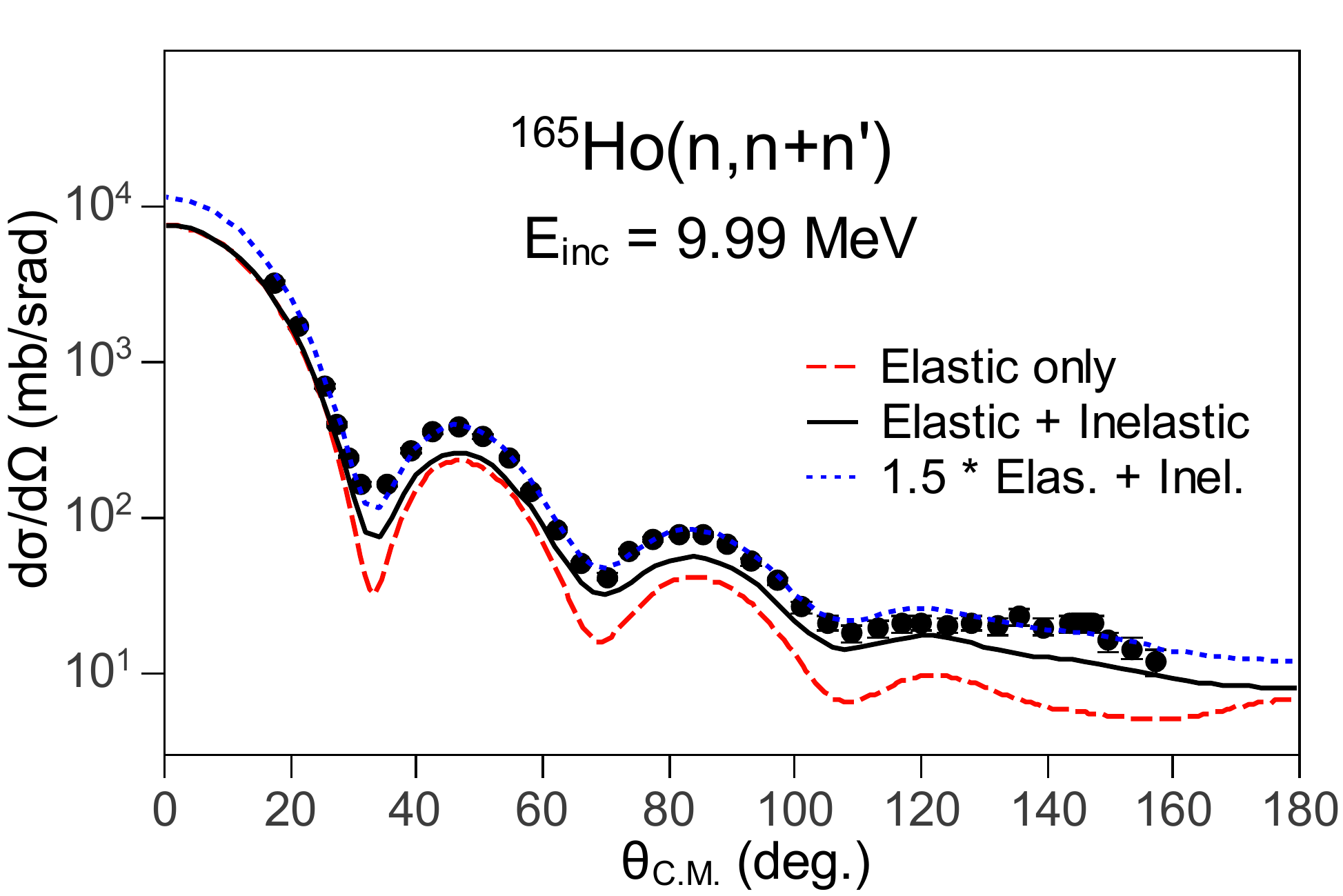}}
\end{center}
\caption{(color online) Comparison between the $^{165}$Ho(n,n+n') angular distributions calculated
using our coupled-channel model and the data set from Ref.~\cite{Smith:2001} for two different
values of incident energies, $E_\mathrm{inc}$ = 4.51 and 9.99 MeV. Experimental data contain
contributions from the elastic and inelastic channels.}
\label{Fig:Ho165Smith}
\end{figure}

The accuracy of our model predictions  was also tested by comparing with the more
recent experimental results  of Ref.~\cite{Smith:2001}. In that work, new measurements
of angular distributions with unresolved contributions from elastic and inelastic channels
for $^{165}$Ho are presented for incident energies $E_\mathrm{inc}$ ranging from 4.51
to 9.99~MeV. In Fig.~\ref{Fig:Ho165Smith} we present such comparisons for the lowest (top panel)
and highest (bottom panel) incident energies of the set. We present in Fig.~\ref{Fig:Ho165Smith},
for comparison purposes, the calculated angular distributions for only the elastic channel as
red-dashed curves, while the solid-black curves correspond to calculations with the inelastic
contributions added up to the elastic one.   In this case, it is seen that, even after adding
the contributions from the inelastic states, our predictions consistently fall slightly below 
the data points. As an indication that our model is reproducing very well the
measured shape of angular distributions, we also plot in Fig.~\ref{Fig:Ho165Smith}
the same elastic-plus-inelastic calculations, but multiplied by a factor of 1.5 blue-dotted curves. This leads to a nearly perfect agreement with
experimental data. %A discussion of this issue will be presented below.

%It is seen Fig.~\ref{Fig:Ho165-Ferrer} that the predictions of our coupled-channel model for the elastic angular distribution (green curve) lies consistently below the experimental data. However, when the calculated contribution from the first inelastic state, which is a 9/2$^-$ state (excitation energy of 94.7~keV), is added (blue curve),  the coupled-channel prediction approaches the observed cross sections. When the second inelastic state (11/2$^-$ state lying at 209.8 keV) is further added  (black curve), we achieve a very good description of the observed quasi-elastic angular distribution.

\begin{figure}
 \begin{center}
\includegraphics[width=.48\textwidth,clip, trim= 9mm 77mm 3mm 62mm]{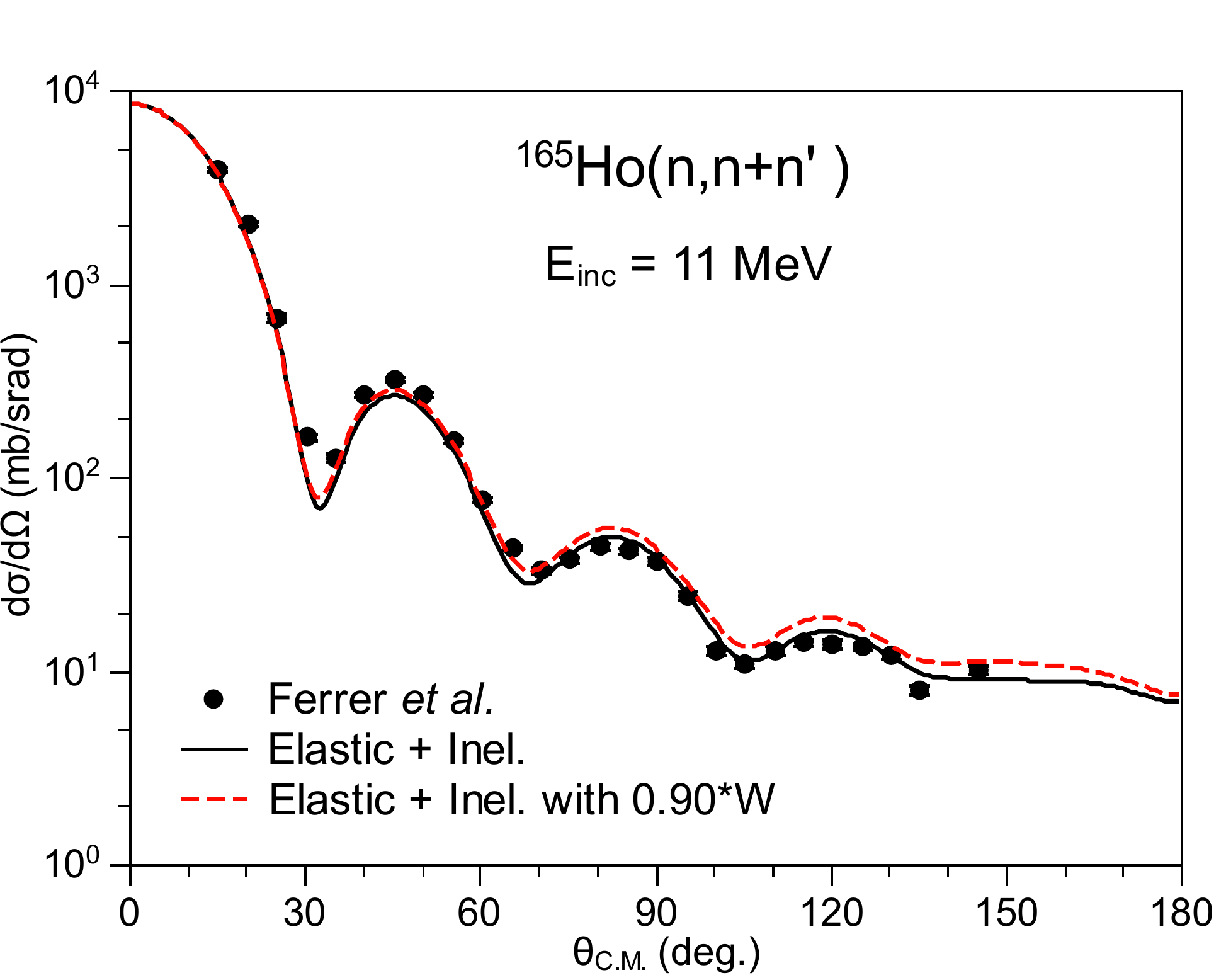}
 \end{center}
  \caption{(color online) Angular distribution for the neutron-induced reaction on
  $^{165}$Ho, at $E_\mathrm{inc}$ = 11.0 MeV. The solid black curve corresponds to
  the results for the elastic channel summed with the inelastic channels. The red-dashed
  curve corresponds to the same calculation but with reduced imaginary potential.
  Experimental data taken from  Ref.~\cite{Ferrer:1977}.}
  \label{Fig:Ho165-Ferrer}
\end{figure}

%As another example,
Fig.~\ref{Fig:Ho165-Ferrer}  presents the predictions of
our model when attempting to describe the angular distribution data for
$^{165}$Ho, at the neutron incident energy of 11 MeV, as measured by
Ferrer \emph{et al.} \cite{Ferrer:1977}. An examination of the experimental conditions of Ref.~\cite{Ferrer:1977}
indicates that in that experiment it was not possible to separate the elastic channel
from the inelastic ones. Therefore, the data points in Fig.~\ref{Fig:Ho165-Ferrer}
should contain inelastic contributions.  For this reason, we plot in Fig.~\ref{Fig:Ho165-Ferrer}
the differential cross sections corresponding to the sum of the elastic and inelastic contributions,
as predicted by our coupled-channel model. For the following discussion we also plot, as the red-dashed curve, the same calculation but with the imaginary components (both volume and surface) of the KD optical potential reduced by
10\%.

The fact that our model describes very well the observed data from the lower and
higher ends of the incident-energy spectrum, as seen in Figs.~\ref{Fig:Ho165-Wagner}
and \ref{Fig:Ho165-Ferrer}, respectively, and also data in between (Fig.~\ref{Fig:Ho165Meadows}),
except for one particular data set  for which the agreement is not as good
(Fig.~\ref{Fig:Ho165Smith}) suggests that there might be an inconsistency between the
experiments from Refs.~\cite{Wagner:1965,Ferrer:1977,Meadows:1971} and the one from
Ref.~\cite{Smith:2001}. The most striking inconsistency is that between the $\approx$10~MeV results of Fig.~\ref{Fig:Ho165Smith} and the 11~MeV results of Fig.~\ref{Fig:Ho165-Ferrer}, since the optical potential is expected to vary slowly and smoothly over this small interval.  We also note that a 10\% reduction in the imaginary potential strength, whose effect is shown at 11~MeV in Fig.~\ref{Fig:Ho165-Ferrer}, would be insufficient to bring the 10-MeV calculations and experiment of Fig.~\ref{Fig:Ho165Smith} into agreement.  At present the source of these discrepancies is not understood.

Considering the simplicity of the model assumptions and the lack of fitted parameters,
we regard the agreement of the predictions of our model with experimental data
as satisfactory.

\subsubsection{Tungsten isotopes}

We analyzed the accuracy of our model when describing the observed angular distributions
of neutrons scattered by the three most abundant tungsten isotopes: $^{182}$W, $^{184}$W,
and $^{186}$W. For this we compared our calculations with experimental data available in the literature.
Guenther \emph{et al.} \cite{Guenther:1982} have measured angular distributions for the elastic and inelastic
(associated with the first 2$^+$ and 4$^+$ excited states) channels for several incident neutron energies. A  measurement of elastic and inelastic angular distributions at $E_{\mathrm{inc}}$=3.4 MeV was made by
Delaroche \emph{et al.} \cite{Delaroche:1981}, while Ref.~\cite{Annand:1985} presents the
results from Annand and Finlay for differential cross section data for  $^{182}$W and $^{184}$W
at $E_{\mathrm{inc}}$=4.87 and 6.0 MeV. The latter experiment also resolved the angular distribution
corresponding to the first 6$^+$ states of the two lighter even-even isotopes.

\begin{figure*}
\begin{center}
   \subfigure{
    \includegraphics[height=.41\textheight,,clip, trim= 9mm 34mm 4mm 29mm]{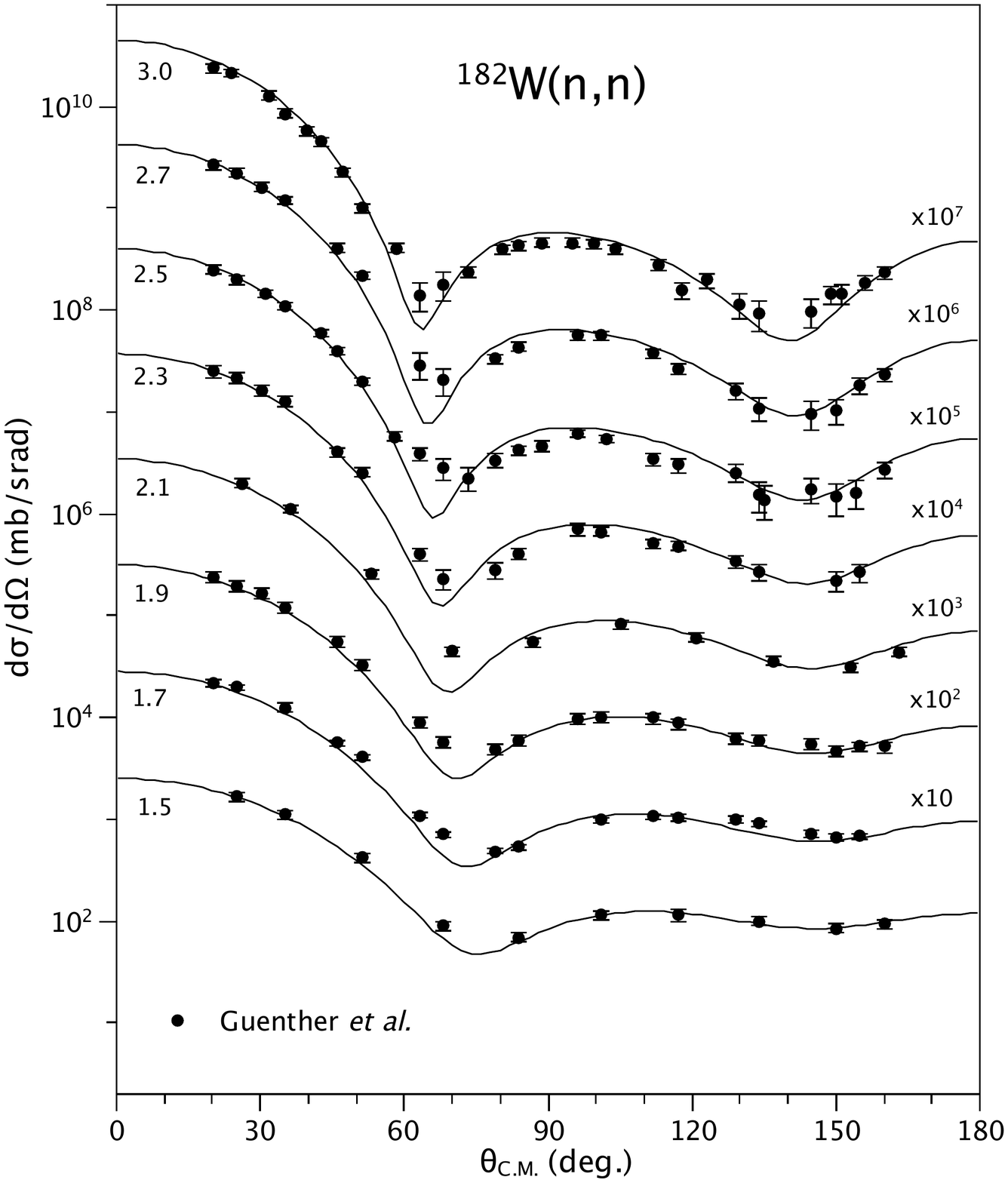}
    \includegraphics[height=.41\textheight,,clip, trim= 15mm 34mm 4mm 29mm]{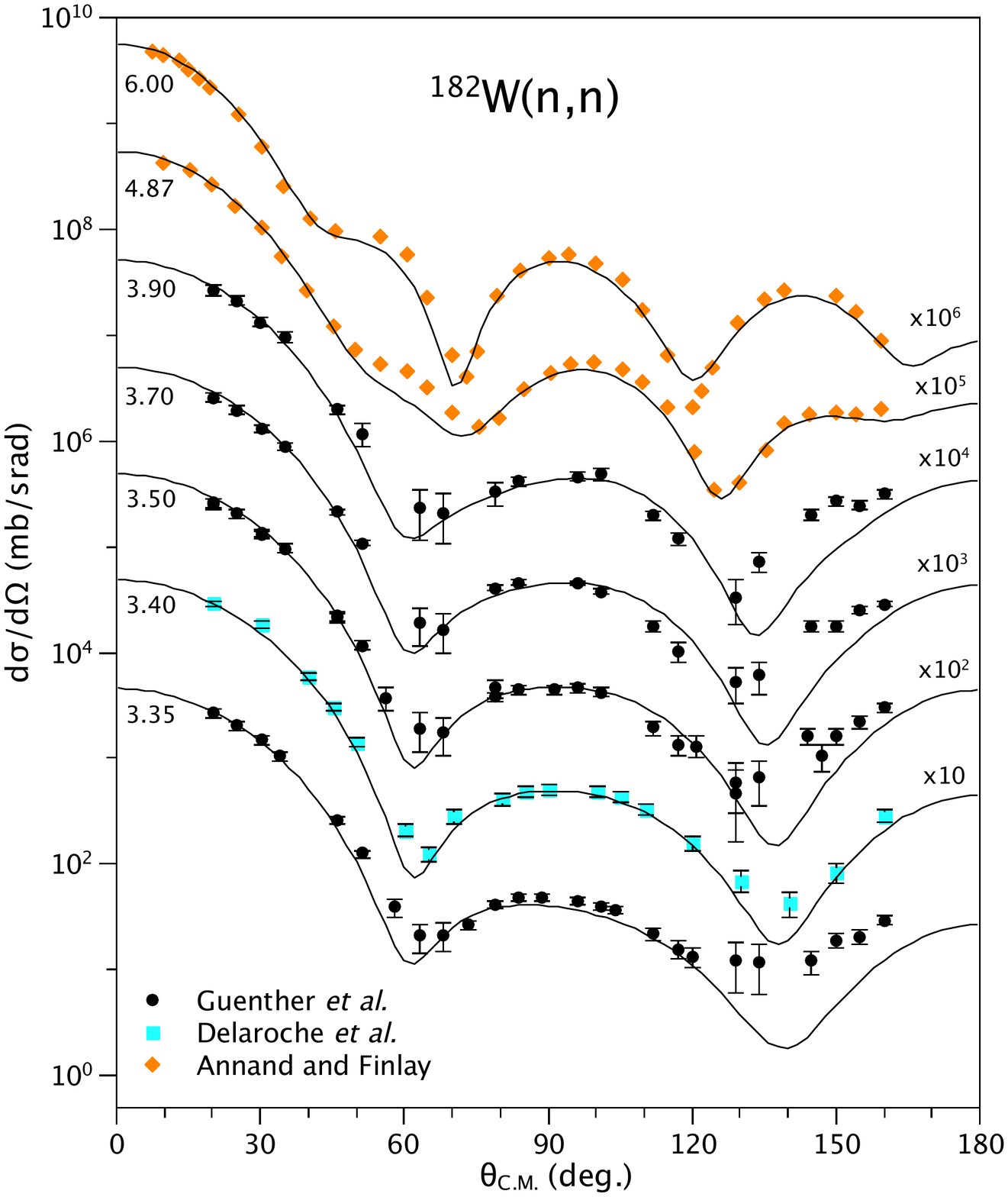}
    }
%   \subfigure{ \includegraphics[height=.41\textheight,,clip, trim= 15mm 34mm 4mm 29mm]{w182-Elas-2}}
  %\includegraphics[height=.5\textheight,,clip, trim= 38mm 22mm 34mm 19mm]{gd158-Bauge}
\end{center}
\caption{(color online) Elastic angular distributions for neutron-induced reactions on $^{\mathrm{182}}$W.
The curves correspond to predictions by our coupled-channel model. Numbers on the left-hand side of
each plot indicate, in MeV, the values of incident energy at which the cross sections were measured,
while the numbers on the right-hand side correspond to the multiplicative factor applied to be able to
plot data from different incident energies in the same graph. Experimental data taken from
Refs.~\cite{Guenther:1982,Delaroche:1981,Annand:1985} and their correspondence to each data set is
indicated in the legends.}
\label{Fig:W182-Elas}
\end{figure*}

Fig.~\ref{Fig:W182-Elas} shows the predictions of our model for the elastic angular
distributions in the case of $^{182}$W when compared to observed experimental data.
Apart from some discrepancies observed at the backward angles for some incident energies
in the region $3.35\leqslant E_{\mathrm{inc}} \leqslant 3.90$ MeV, it can be seen in
Fig.~\ref{Fig:W182-Elas} that excellent agreement is obtained.

\begin{figure*}
\begin{center}
   \subfigure[$^{182}$W 2$^{+}$ inelastic angular distributions.]{\label{Fig:W182:2+}
   \includegraphics[height=.44\textheight,,clip, trim= 9mm 28mm 4mm 23mm]{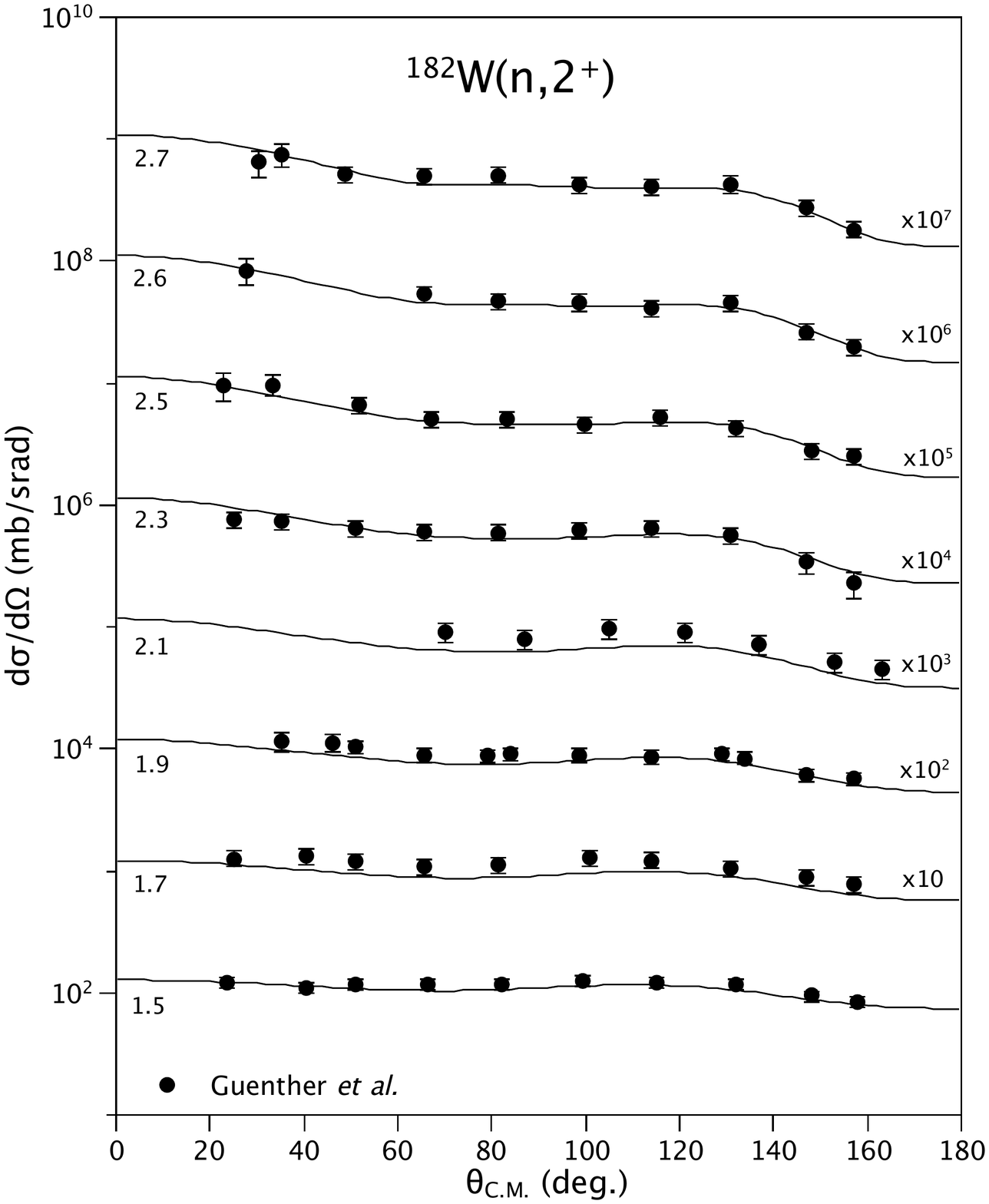}
   \includegraphics[height=.44\textheight,,clip, trim= 15mm 28mm 4mm 23mm]{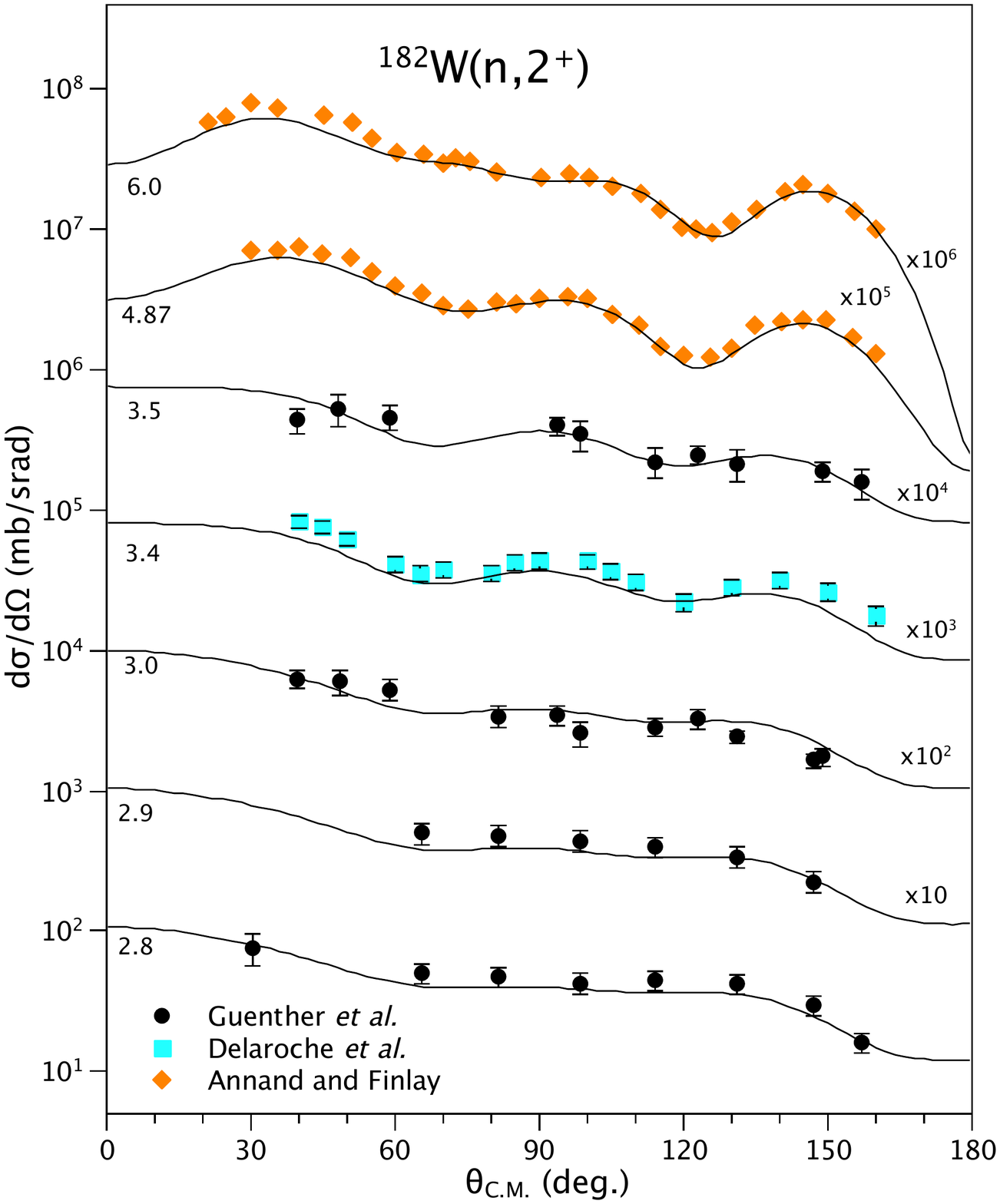}}

   \subfigure[$^{182}$W 4$^{+}$ inelastic angular distributions.]{\label{Fig:W182:4+}
  \includegraphics[height=.35\textheight,,clip, trim= 9mm 55mm 4mm 45mm]{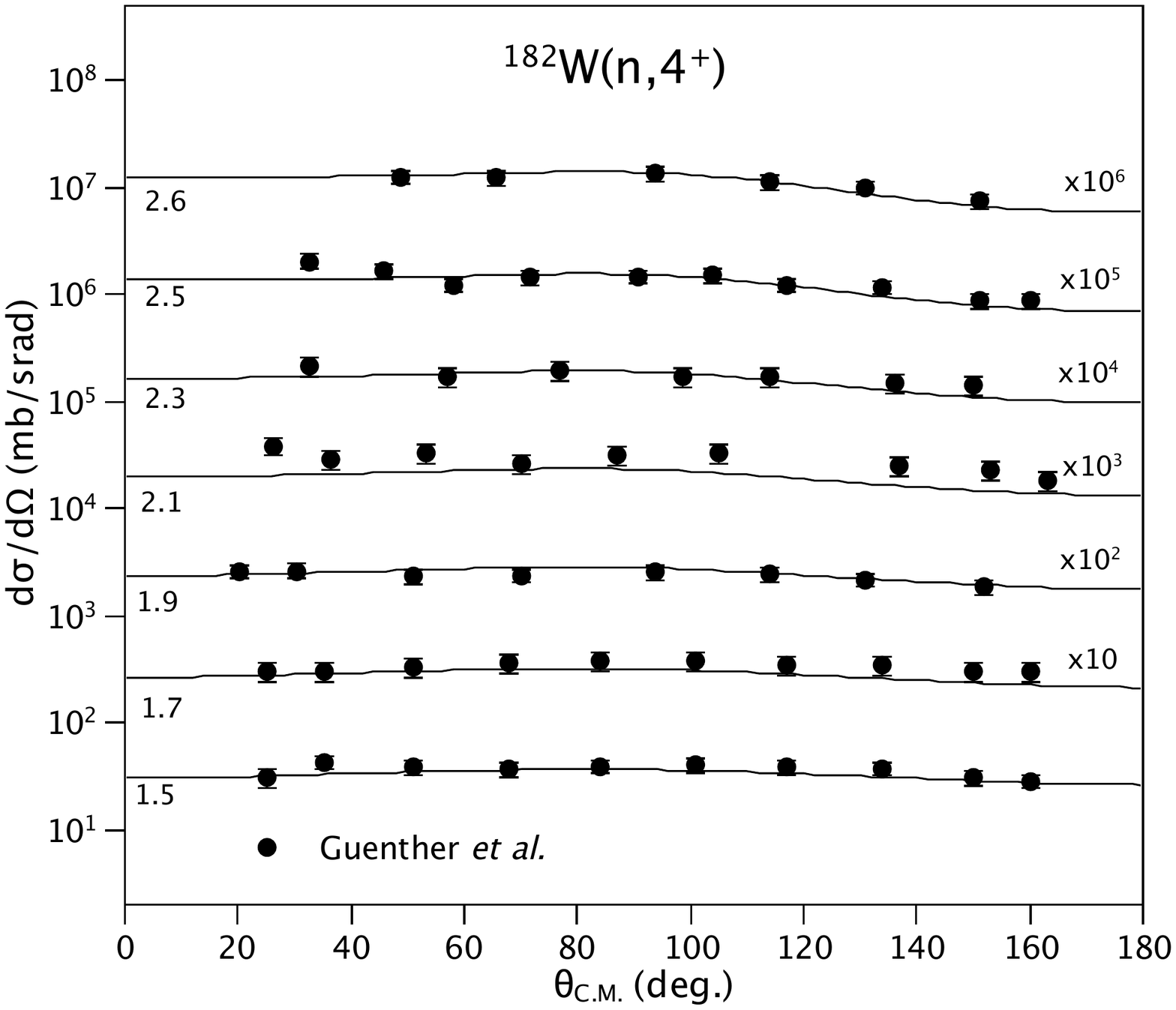}
  \includegraphics[height=.35\textheight,,clip, trim= 15mm 55mm 4mm 45mm]{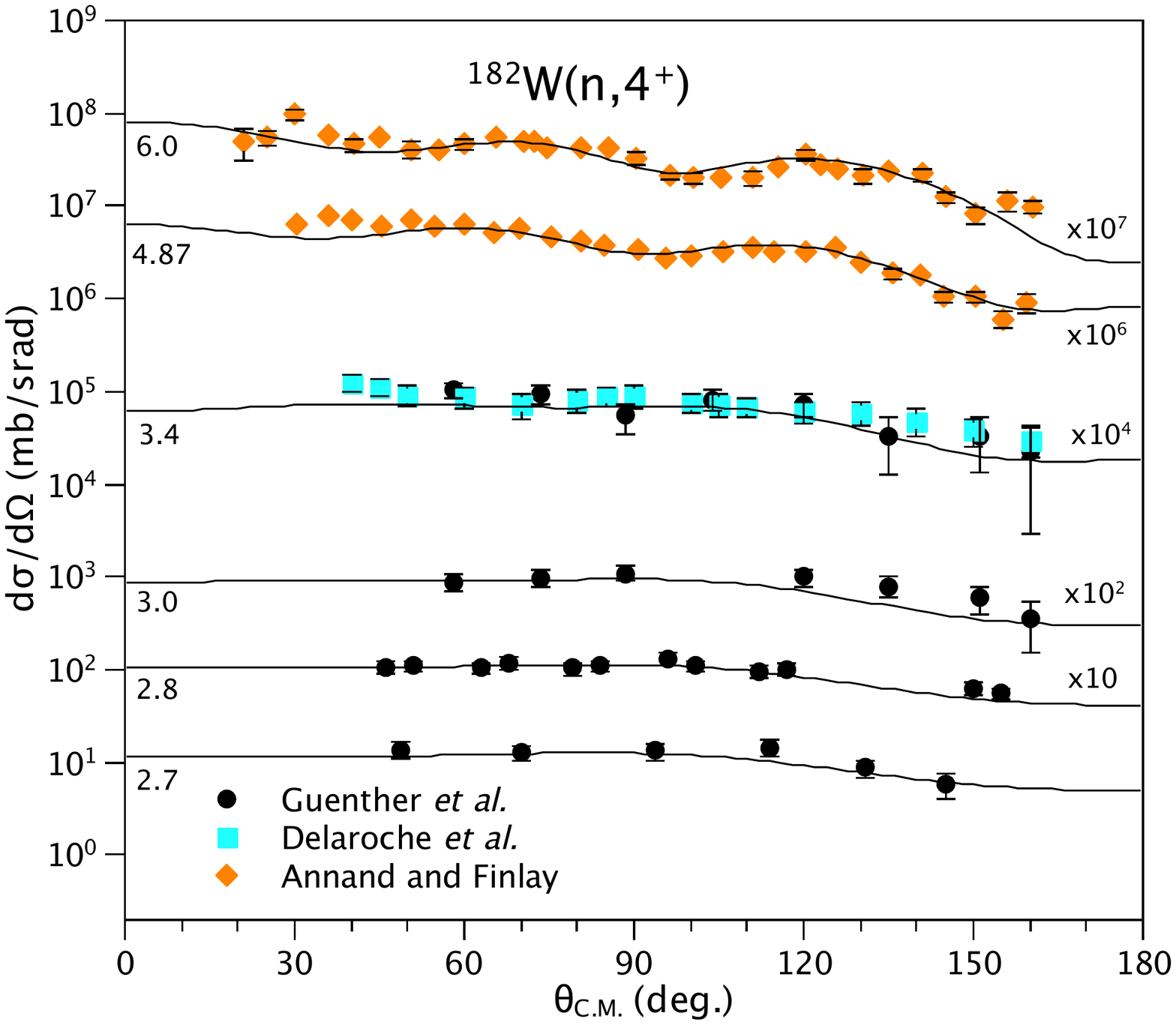}}
 %  \subfigure[$^{182}$W 6$^{+}$ inelastic angular distributions.]{\label{Fig:W182:6+}
  %   \includegraphics[height=.15\textheight, clip, trim= 11mm 103mm 6mm 100mm]{w182-6+}  }
\end{center}
\caption{(color online) Inelastic angular distributions for neutron-induced reactions
on $^{\mathrm{182}}$W.  The curves correspond to predictions by our coupled-channel model.
Numbers on the left-hand side of each plot indicate, in MeV, the values of incident energy
at which the cross sections were measured, while the numbers on the right-hand side correspond
to the multiplicative factor applied to be able to plot data from different incident energies in
the same graph. Experimental data taken from Refs.~\cite{Guenther:1982,Delaroche:1981,Annand:1985}
and their correspondence to each data set is indicated in the legends.}\label{Fig:W182-Inel}
\end{figure*}

In Fig.~\ref{Fig:W182-Inel}, we present the results of calculations using our coupled-channel
model for the inelastic differential cross sections of  $^{182}$W  for the 2$^+$
(Fig.~\ref{Fig:W182:2+}), and 4$^+$ (Fig.~\ref{Fig:W182:4+}) excited states,  which have excitation
energies $E^*$ of 100.1 keV, and 329.4 keV, respectively. The agreement with experimental data is very good,
for both  cases of 2$^+$ and 4$^+$ excitations.

\begin{figure*}
\begin{center}
   \subfigure{
   \includegraphics[height=.41\textheight,,clip, trim= 9mm 34mm 4mm 29mm]{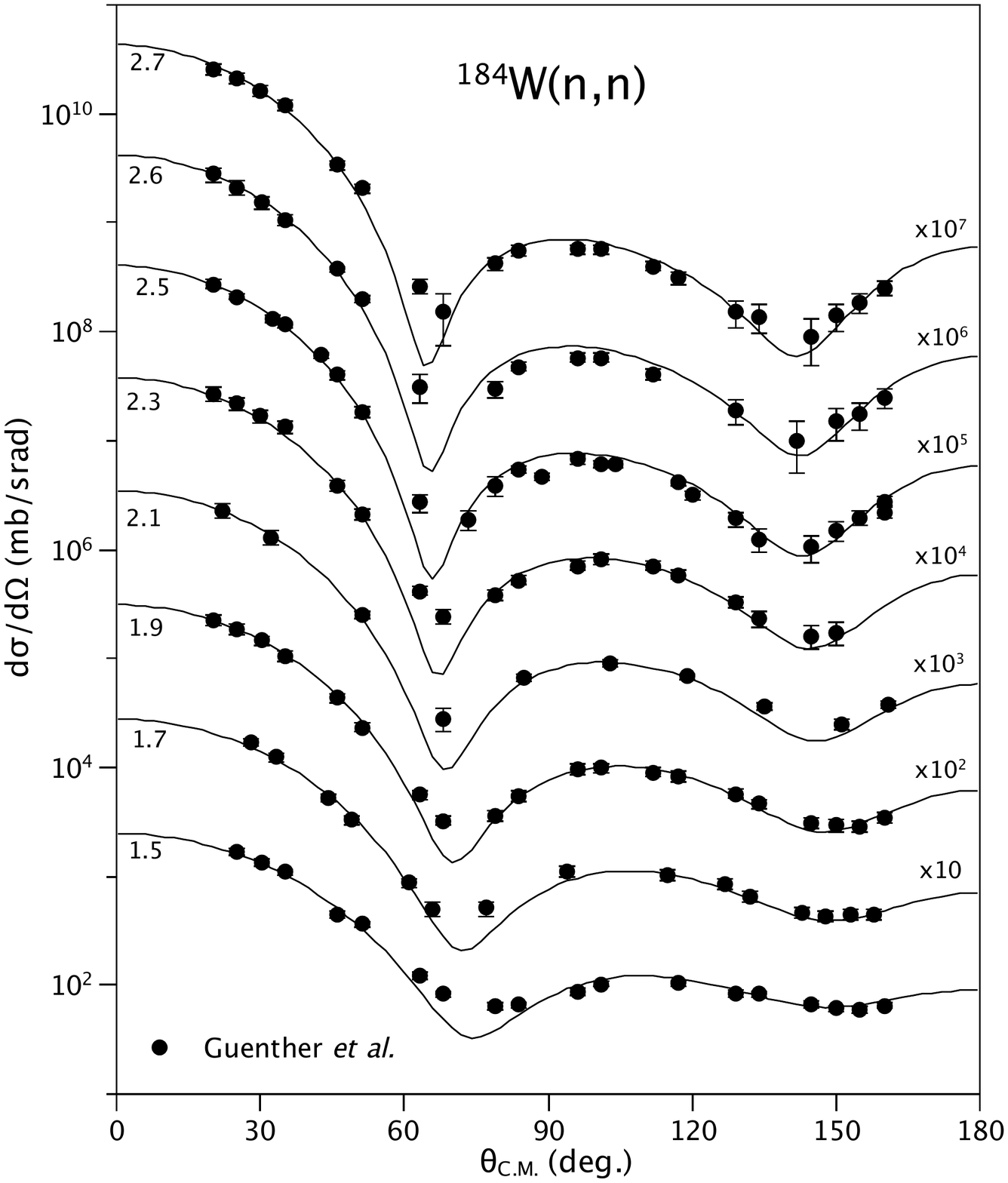}
   \includegraphics[height=.41\textheight,,clip, trim= 15mm 34mm 4mm 29mm]{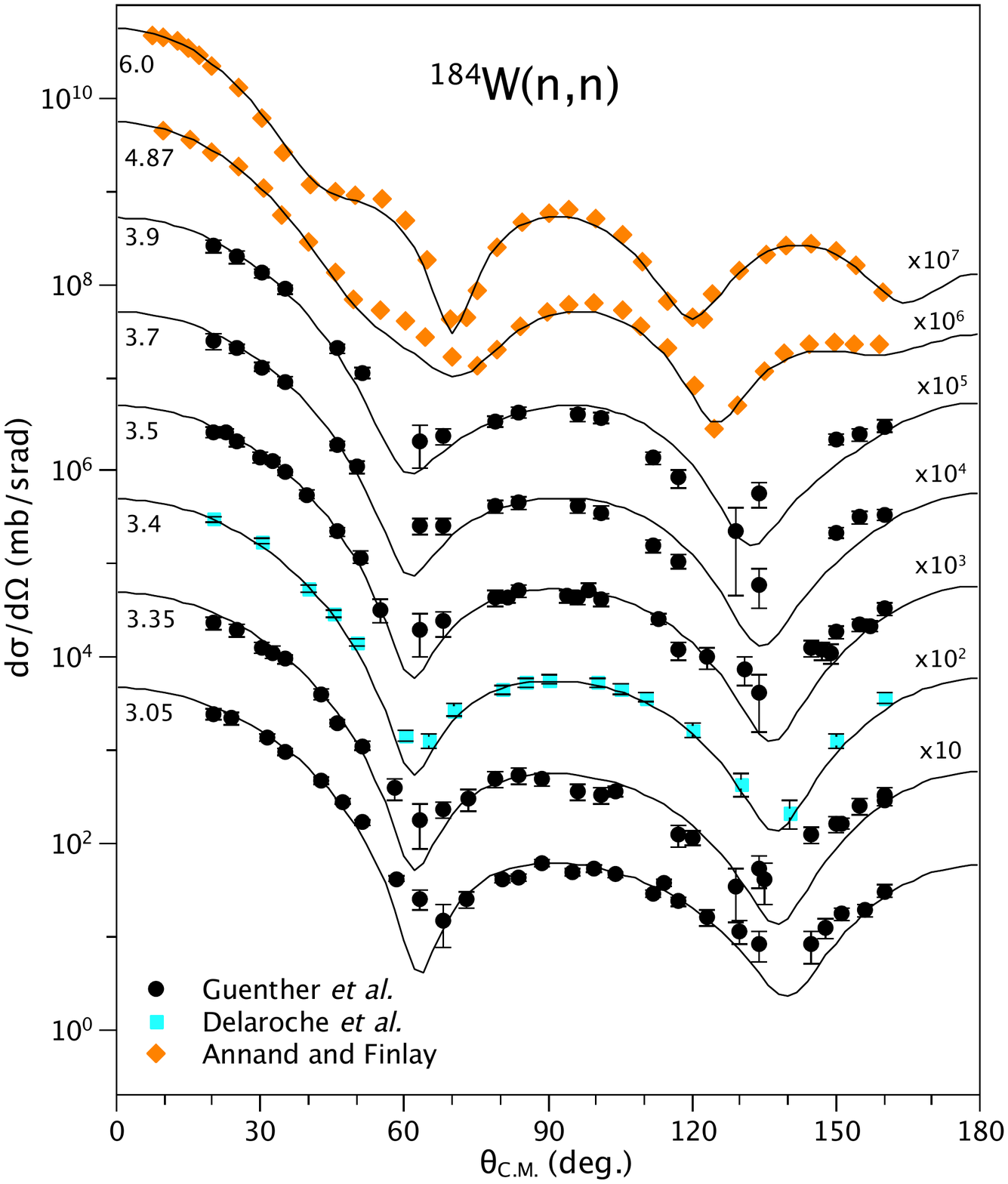}
   }
\end{center}
\caption{(color online) Elastic angular distributions for neutron-induced reactions on $^{\mathrm{184}}$W.
The curves correspond to predictions by our coupled-channel model. Numbers on the left-hand side of each plot
indicate, in MeV, the values of incident energy at which the cross sections were measured, while the numbers
on the right-hand side correspond to the multiplicative factor applied to be able to plot data from different
incident energies in the same graph. Experimental data taken from
Refs.~\cite{Guenther:1982,Delaroche:1981,Annand:1985} and their correspondence to each data set is
indicated in the legends.}
\label{Fig:W184-Elas}
\end{figure*}

We present in Fig.~\ref{Fig:W184-Elas} the results obtained for $^{\mathrm{184}}$W elastic
angular distributions. The agreement in this case is even better than the one obtained for $^{\mathrm{182}}$W,
shown in Fig.~\ref{Fig:W182-Elas}, considering that the discrepancies at backward angles
in the $3.35\leqslant E_{\mathrm{inc}} \leqslant 3.90$ region are much smaller for $^{\mathrm{184}}$W
than for $^{\mathrm{182}}$W.

\begin{figure*}
\begin{center}
   \subfigure[$^{184}$W 2$^{+}$ inelastic angular distributions.]{\label{Fig:W184:2+}
   \includegraphics[height=.34\textheight,,clip, trim= 9mm 48mm 4mm 53mm]{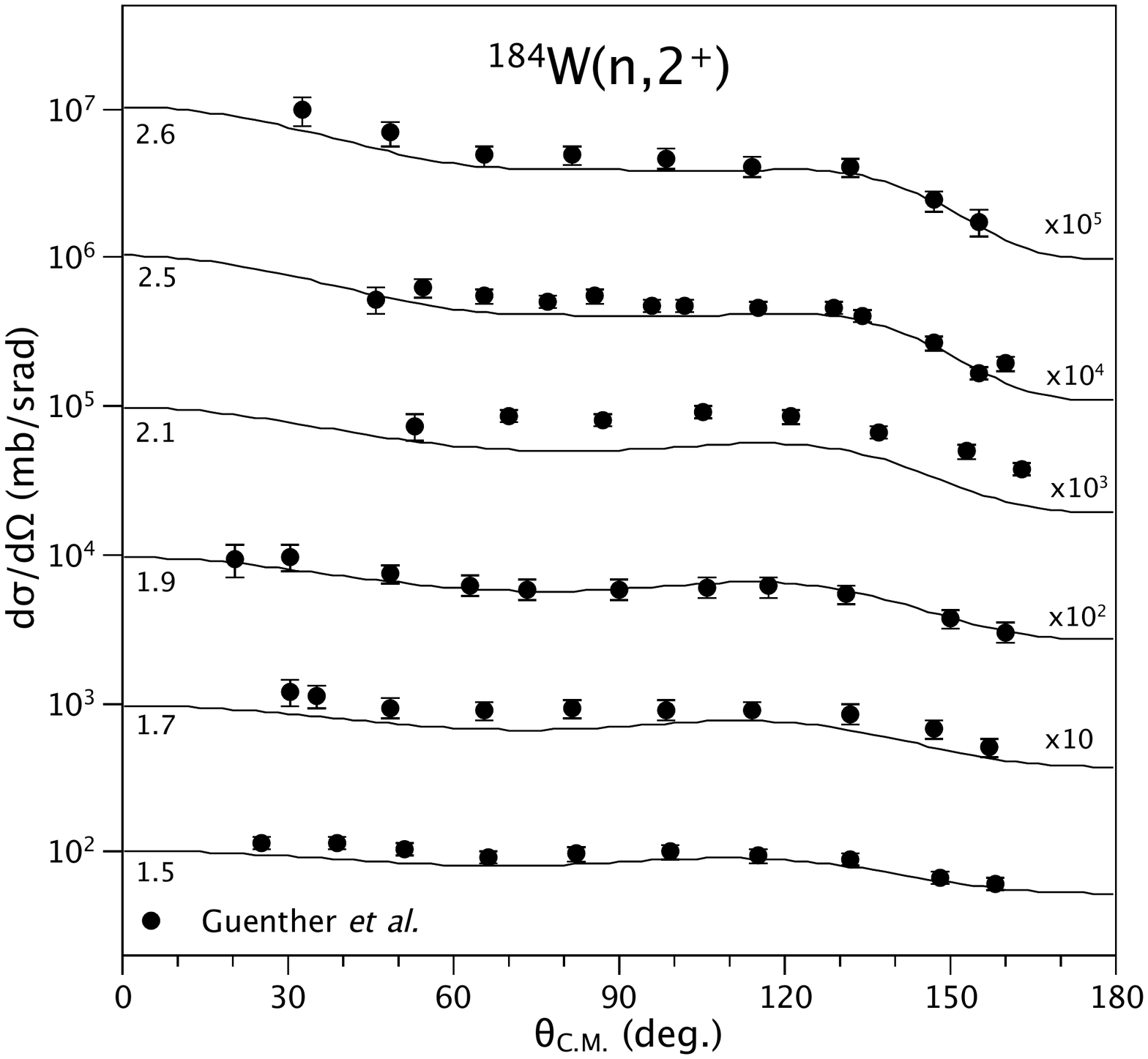}
   \includegraphics[height=.34\textheight,,clip, trim= 15mm 48mm 4mm 53mm]{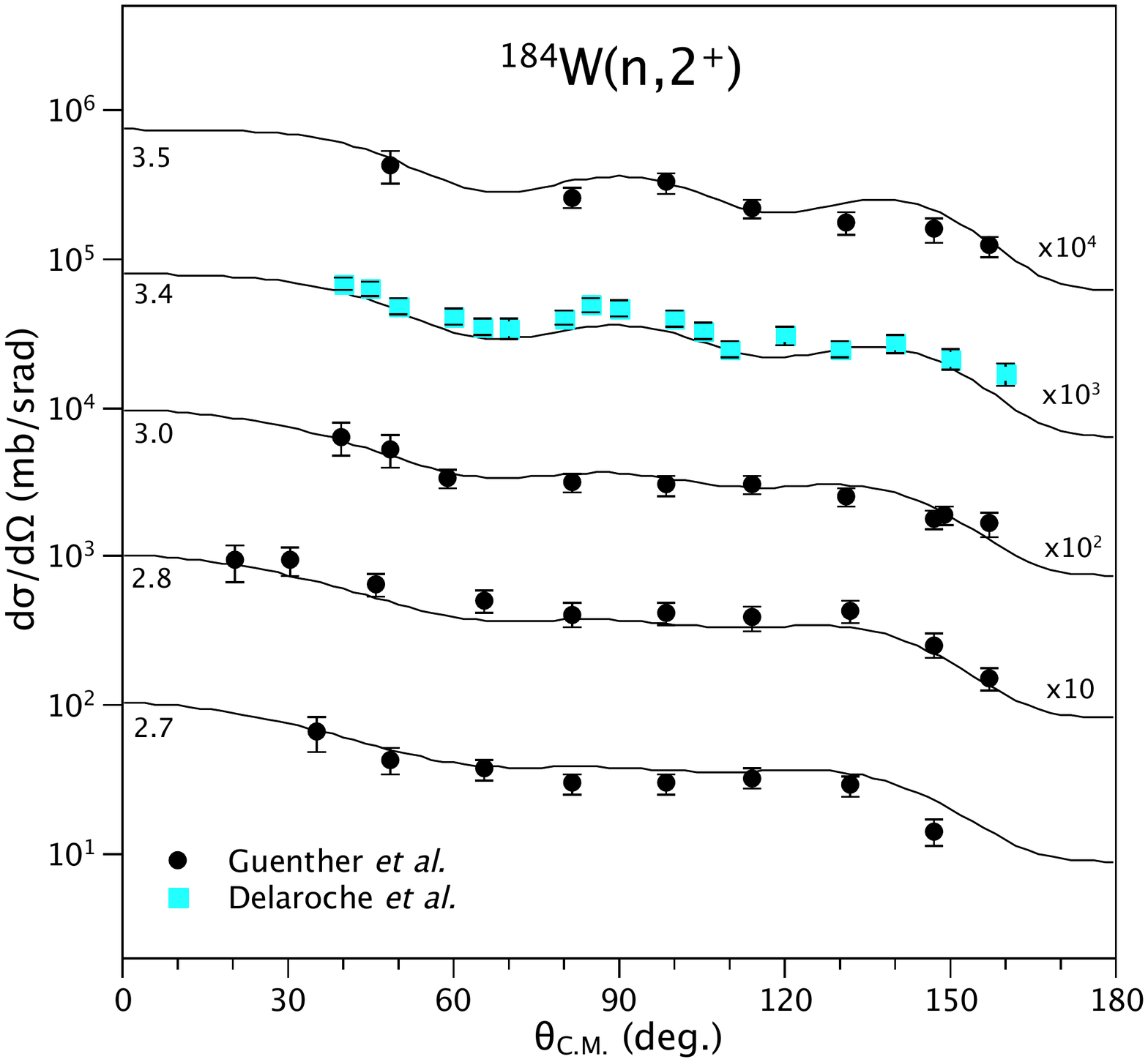}}

   \subfigure[$^{184}$W 4$^{+}$ inelastic angular distributions.]{\label{Fig:W184:4+}
  \includegraphics[height=.30\textheight,,clip, trim= 9mm 59mm 4mm 65mm]{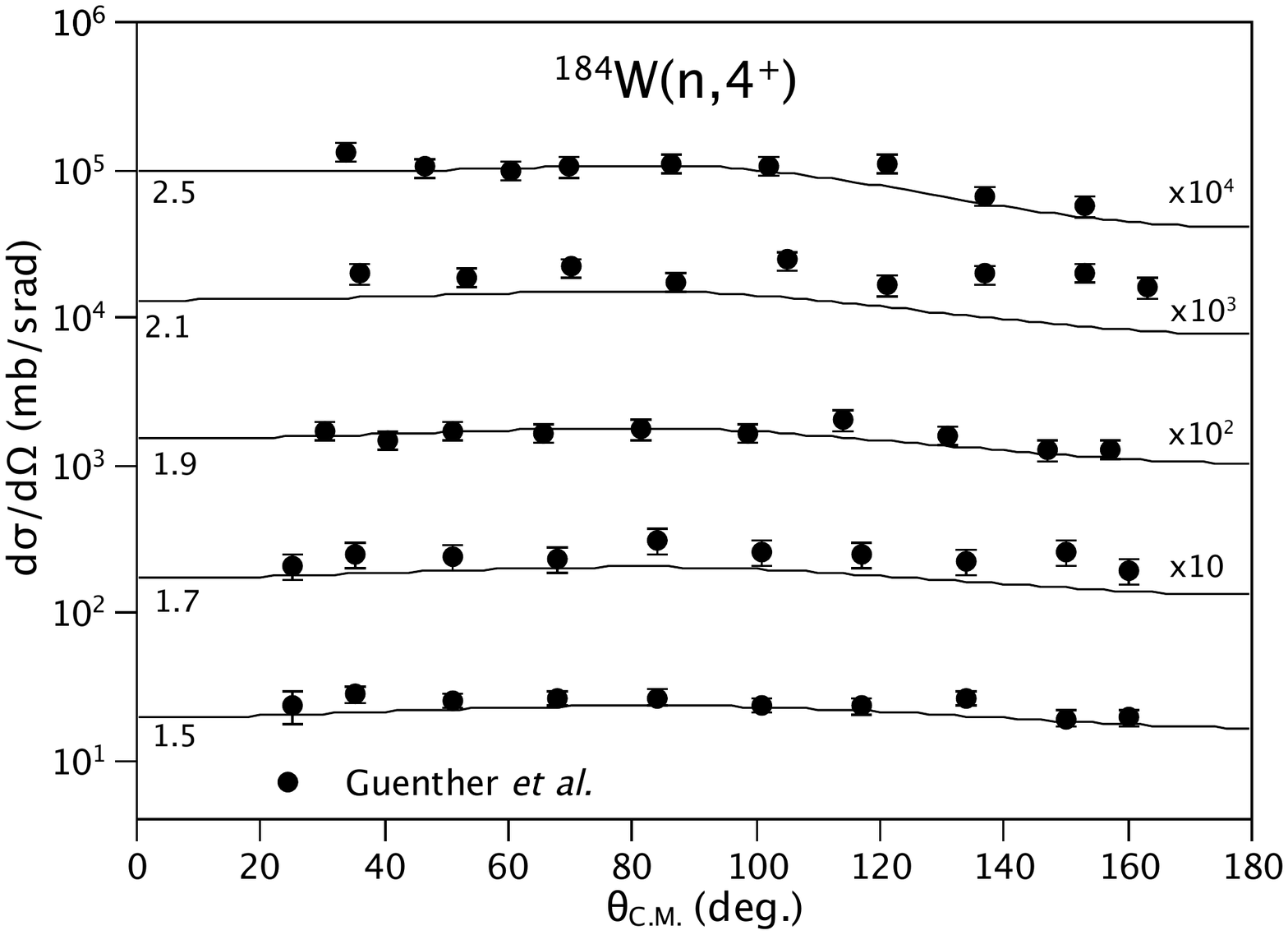}
  \includegraphics[height=.30\textheight,,clip, trim= 15mm 59mm 4mm 65mm]{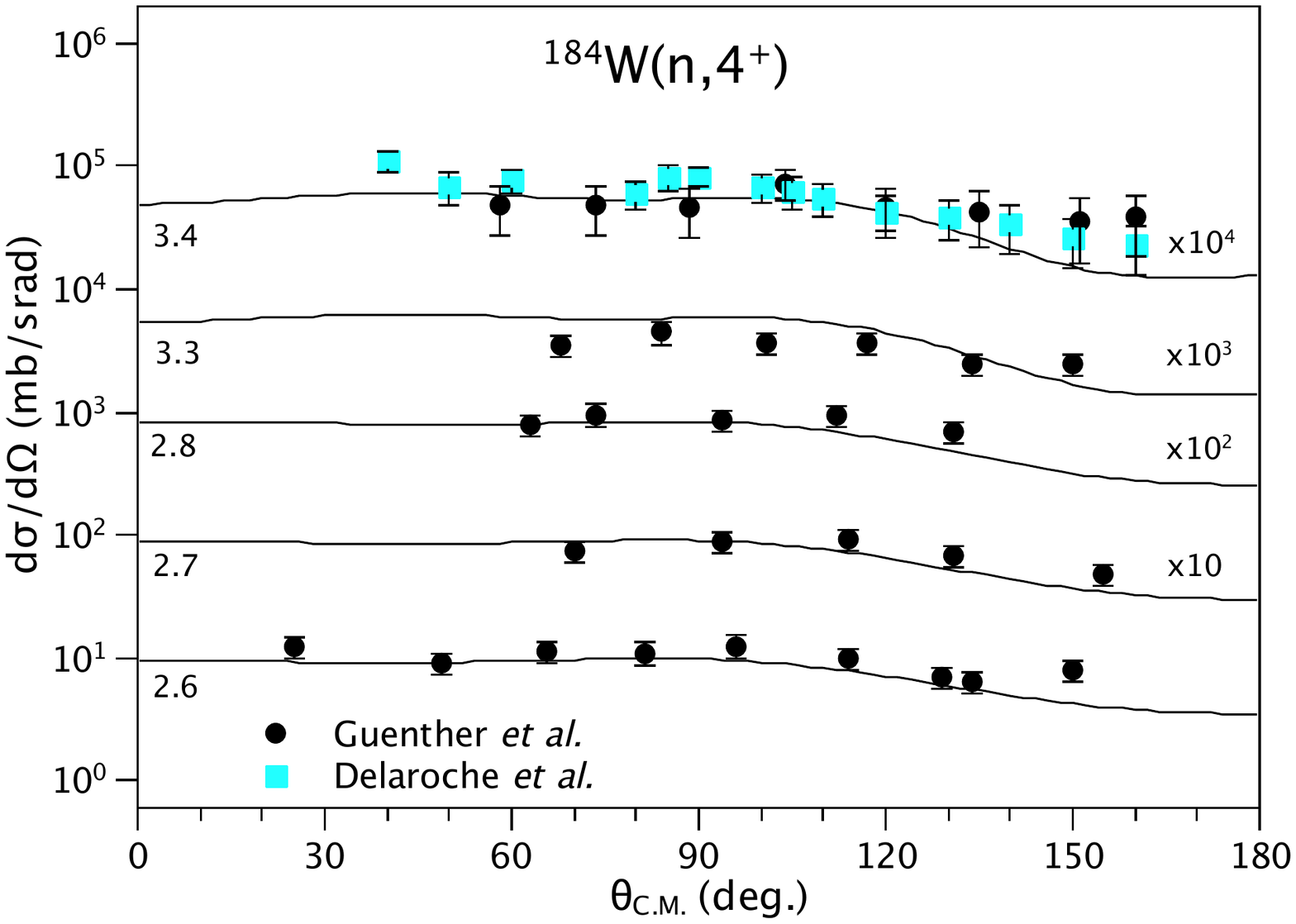}}\\
\end{center}
\caption{(color online) Inelastic angular distributions for neutron-induced reactions on $^{\mathrm{184}}$W.
The curves correspond to predictions by our coupled-channel model. Numbers on the left-hand side of each
plot indicate, in MeV, the values of incident energy at which the cross sections were measured,
while the numbers on the right-hand side correspond to the multiplicative factor applied to be able to
plot data from different incident energies in the same graph. Experimental data taken from
Refs.~\cite{Guenther:1982,Delaroche:1981,Annand:1985} and their correspondence to each data set is
indicated in the legends.}
\label{Fig:W184-Inel}
\end{figure*}

The results for inelastic angular distributions of $^{\mathrm{184}}$W are presented in
Fig.~\ref{Fig:W184-Inel}. Fig.~\ref{Fig:W184:2+} shows the differential cross sections
corresponding to the first 2$^+$ state ($E^{*}=111.2$ keV), while Fig.~\ref{Fig:W184:4+}
presents results for the first 4$^+$ state ($E^{*}=364.1$ keV) of  $^{\mathrm{184}}$W.
Similarly to the case of $^{\mathrm{182}}$W shown in Fig.~\ref{Fig:W182-Inel}, we achieve very good
agreement between experimental differential cross-section data with the ones calculated
via our model for $^{\mathrm{184}}$W. Larger discrepancies are observed only for a particular
incident energy, $E_{\mathrm{inc}}=2.1$ MeV, for both 2$^+$ (Fig.~\ref{Fig:W184:2+}) and
4$^+$ (Fig.~\ref{Fig:W184:4+}) cross sections, where we underestimate the experimental data.

%\begin{figure*}
%\begin{center}
%   \subfigure{ \includegraphics[height=.41\textheight,,clip, trim= 9mm 34mm 4mm 29mm]{w186-Elas-1}  \includegraphics[height=.41\textheight,,clip, trim= 15mm 34mm 4mm 29mm]{w186-Elas-2}}
%\end{center}
%\caption{(color online) Elastic angular distributions for neutron-induced reactions on $^{\mathrm{186}}$W.  The curves correspond to predictions by our coupled-channel model. Numbers on the left-hand side of each plot indicate, in MeV, the values of incident energy at which the cross sections were measured, while the numbers on the right-hand side correspond to the multiplicative factor applied to be able to plot data from different incident energies in the same graph. Experimental data taken from Refs.~\cite{Guenther:1982,Delaroche:1981,Annand:1985} and their correspondence to each data set is indicated in the legends.}
%\label{Fig:W186-Elas}
%\end{figure*}

\begin{figure}
\begin{center}
   \subfigure[$^{182}$W 6$^{+}$ inelastic angular distributions.]{\label{Fig:W182:6+}
     \includegraphics[width=0.48\textwidth, clip, trim= 11mm 103mm 6mm 100mm]{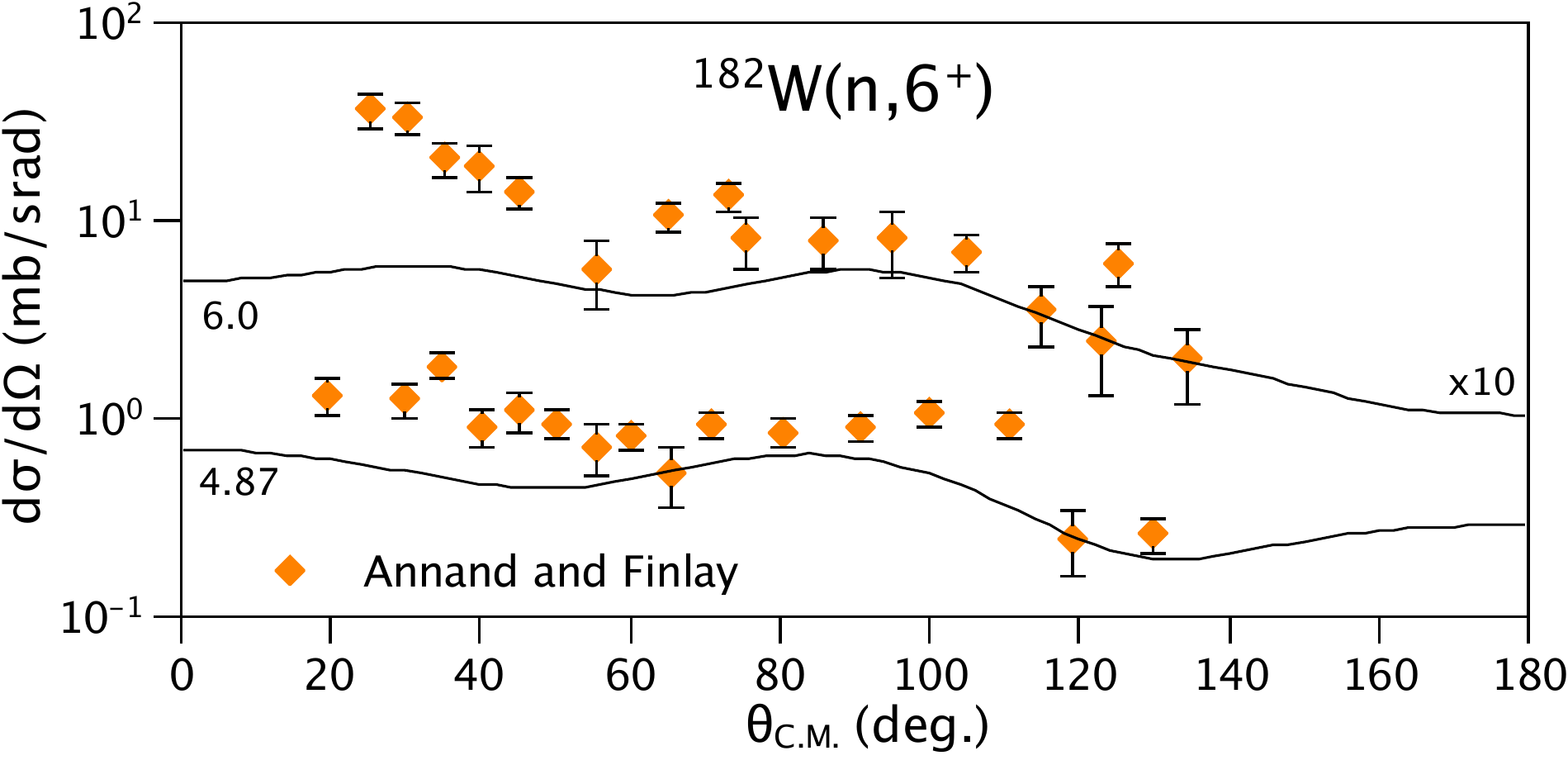}  }
        \subfigure[$^{184}$W 6$^{+}$ inelastic angular distributions.]{\label{Fig:W184:6+}
     \includegraphics[width=0.48\textwidth, clip, trim= 11mm 103mm 6mm 100mm]{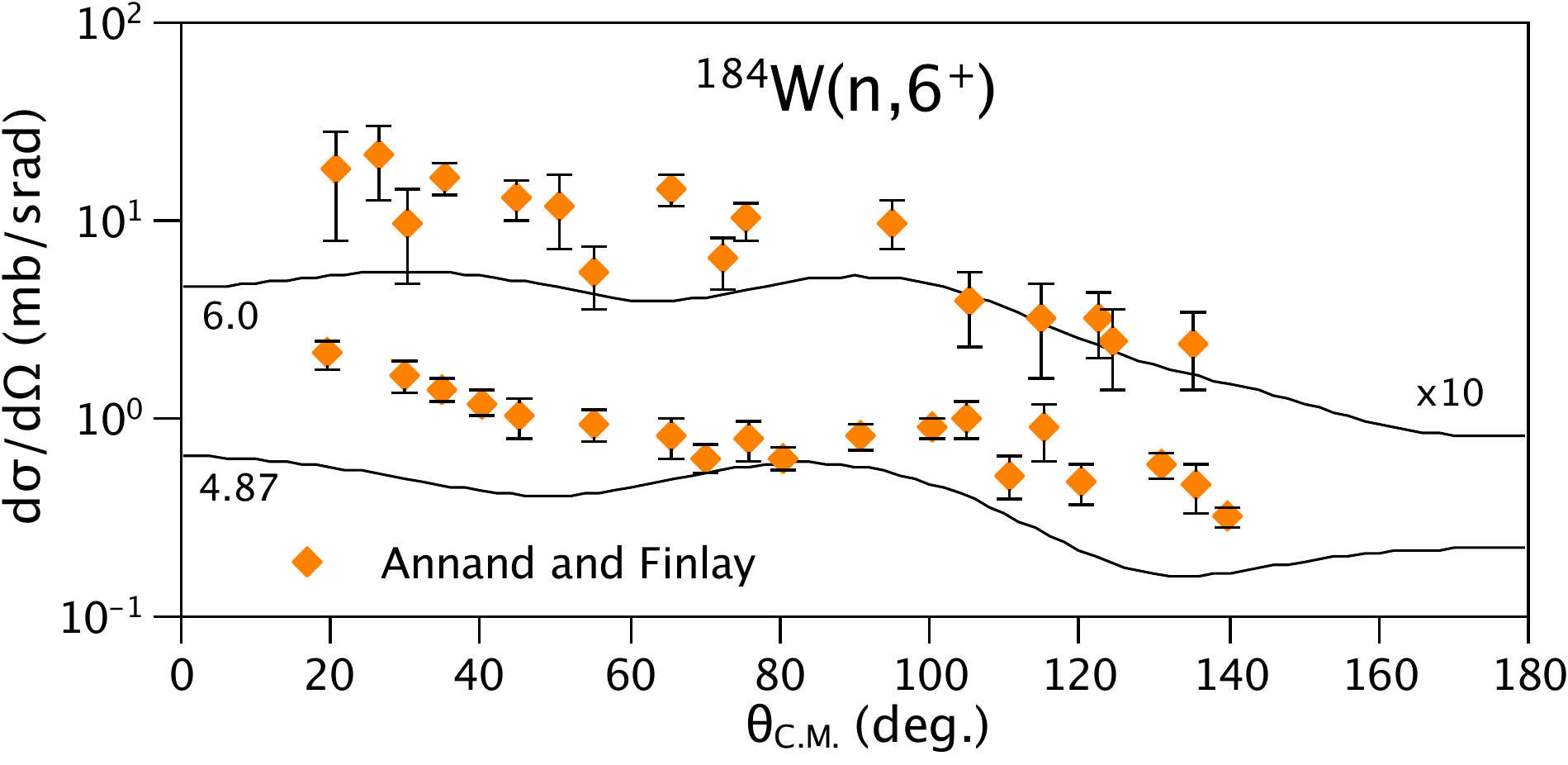}  }

\end{center}
\caption{(color online) Inelastic angular distributions for neutron-induced reactions on
$^{\mathrm{182}}$W.  The curves correspond to predictions by our coupled-channel model. Numbers
on the left-hand side of each plot indicate, in MeV, the values of incident energy at which the
cross sections were measured, while the numbers on the right-hand side correspond to the multiplicative
factor applied to be able to plot data from different incident energies in the same graph. Experimental
data taken from Refs.~\cite{Guenther:1982,Delaroche:1981,Annand:1985} and their correspondence to each data
set is indicated in the legends.}
\label{Fig:W:6+}
\end{figure}

In Fig.~\ref{Fig:W:6+}, we present the results of calculations within our coupled-channel model
for the inelastic differential cross sections for the  6$^+$ state for both $^{182}$W
(Fig.~\ref{Fig:W182:6+}) and $^{184}$W (Fig.~\ref{Fig:W184:6+}) neutron-induced reactions.
The excitation energies of such states are $E^*$ = 680.5 keV and 748.3 keV, respectively.
In the case of the 6$^+$ angular distributions, the agreement with experimental data is not as good as
in the cases of the 2$^+$ and 4$^+$ channels (Figs. \ref{Fig:W182-Inel} and \ref{Fig:W184-Inel}).
This is not surprising since our calculations do not include a direct excitation of the 6$^+$ state 
via off-diagonal elements connecting the elastic channel and the 6$^+$ state in the coupling potential. 
%(which usually depends on the $\beta_6$ deformation parameter).

We have also performed calculations for the elastic and inelastic (2$^+$ and 4$^+$ channels) angular
distributions of neutrons scattered by $^{\mathrm{186}}$W nuclide. We do not present them here, as those
results are very similar to the ones obtained for $^{\mathrm{182,184}}$W (Figs. \ref{Fig:W182-Elas}-\ref{Fig:W184-Inel}).

%\begin{figure*}
 %\begin{center}
% \begin{minipage}[c]{.35\textwidth}
 % \includegraphics[height=.420\textheight, clip, trim= 5mm -0mm -2mm -2mm]{graph-W186-Elas}
 % \end{minipage}  \hspace{5mm}
  % \begin{minipage}[c]{.55\textwidth}
 % \includegraphics[scale=.5, clip, trim= 14mm 1mm 0mm 0mm]{graph-W186-Inel2+}  \\ \vspace{-10.5cm}
 % \includegraphics[scale=.5, clip, trim= 14mm 1mm 0mm 0mm]{graph-W186-Inel4+}
  %  \end{minipage}
% \end{center}
% \vspace{-5.0cm}
 % \caption{Picture to fixed height}
%\end{figure*}

\section{Summary and conclusions}
\label{Sec:Conclusion}

In this paper we presented extensive results for a coupled-channel model designed to accurately predict
differential and integral cross sections of neutron-induced reactions on statically-deformed nuclei, such as
those of the rare-earth region. The method consists in statically deforming a spherical optical potential that
well describes non-deformed nuclei in the neighboring regions, and then carrying out coupled-channels calculations using a sufficient number of the rotational excited states of
the ground state band to achieve convergence. In this particular work we adopted the spherical global Koning-Delaroche \cite{KD} optical model potential (OMP). We leave the OMP unmodified, except for a small correction in the radii to ensure nuclear volume conservation. The idea behind this model is that, due to the low-lying values of excitation energy and consequent near validity of the adiabatic approximation, we can explicitly treat the degrees of freedom associated with strong deformation while all other degrees of freedom are
accounted for by the spherical OMP with its parameters unmodified.

We applied our model to nuclear reactions having $^{\mathrm{158,160,Nat}}$Gd, $^{\mathrm{165}}$Ho,
and $^{\mathrm{182,184,186}}$W as targets. Comparison of our calculations with experimental data indicated remarkable agreement, 
for both total cross sections and angular distributions. Even though the
agreement could be further improved by fitting OMP parameters for the individual nuclei, the importance 
of the present model lies in the achievement of such results \emph{without} any parameter adjustment, 
using as input only the potential and the experimental quadrupole and hexadecupole deformation parameters. 
Therefore, the conclusions presented here could reliably be extrapolated to other statically-deformed rare-earth 
nuclei for which little or no experimental data is available.

%\begin{theacknowledgments}
\section{Acknowledgments}
%\end{theacknowledgments}

The work at Brookhaven National Laboratory was sponsored by the Office of Nuclear
Physics, Office of Science of the U.S. Department of
Energy under Contract No. DE-AC02-98CH10886 with
Brookhaven Science Associates, LLC.

\bibliographystyle{apsrev4-1} % temporarily commented out by fsd
\bibliography{cc_PRC}

\end{document}